\documentclass[12pt,preprint]{aastex}

\begin{document}

\title{On the Disruption of Star Clusters in a Hierarchical Interstellar Medium}

\author{Bruce G. Elmegreen}
\affil{IBM T. J. Watson Research Center, 1101 Kitchawan Road, Yorktown
Heights, New York 10598 USA} \email{bge@us.ibm.com}

\author{Deidre A. Hunter}
\affil{Lowell Observatory, 1400 West Mars Hill Road, Flagstaff, Arizona
86001 USA} \email{dah@lowell.edu}

\begin{abstract}
The distribution of the number of clusters as a function of mass $M$
and age $T$ suggests that clusters get eroded or dispersed in a regular
way over time, such that the cluster number decreases inversely as an
approximate power law with $T$ within each fixed interval of $M$. This
power law is inconsistent with standard dispersal mechanisms such as
cluster evaporation and cloud collisions. In the conventional
interpretation, it requires the unlikely situation where diverse
mechanisms stitch together over time in a way that is independent of
environment or $M$. Here we consider another model in which the large
scale distribution of gas in each star-forming region plays an
important role. We note that star clusters form with positional and
temporal correlations in giant cloud complexes, and suggest that these
complexes dominate the tidal force and collisional influence on a
cluster during its first several hundred million years. Because the
cloud complex density decreases regularly with position from the
cluster birth site, the harassment and collision rates between the
cluster and the cloud pieces decrease regularly with age as the cluster
drifts. This decrease is typically a power law of the form required to
explain the mass-age distribution. We reproduce this distribution for a
variety of cases, including rapid disruption, slow erosion,
combinations of these two, cluster-cloud collisions, cluster disruption
by hierarchical disassembly, and partial cluster disruption. We also
consider apparent cluster mass loss by fading below the surface
brightness limit of a survey. In all cases, the observed $\log M - \log
T$ diagram can be reproduced under reasonable assumptions.
\end{abstract}

\section{Introduction}

For a galaxy-wide population of star clusters, the distribution of
cluster mass $M$ and age $T$ on a $\log M$ versus $\log T$ diagram is
approximately uniform or slowly varying with age for each fixed range
of $M$ above the detection limit. This slow variation appears in the
density of points on such a diagram, which often has no obvious
gradient along horizontal ($\log T$ axis) lines, or only a small
increase with $\log T$. Clusters have this $M-T$ distribution in the
Large and Small Magellanic Clouds (Hunter, et al. 2003; de Grijs \&
Anders 2006; Chandar, Fall, Whitmore 2006; Chandar et al. 2009),
several dwarf galaxies (e.g., IC 1613, DDO 50, NGC 2366 in Melena et
al. 2009), five spiral galaxies studied by Mora et al. (2009), and the
Antennae (Fall, Chandar \& Whitmore 2005; Whitmore, Chandar, \& Fall
2007).  Local Milky Way clusters have the same mass-age distribution,
as shown by the constancy of the number per unit logarithmic age
interval for a uniform average mass in Figure 3 of Lada \& Lada (2003).
Rafelski \& Zaritsky (2005) also showed this age distribution for the
LMC in their figure 12, which has a measurable slope of -1.1 in a plot
of cluster number per unit age versus log of the age.

A uniform $\log T$ distribution on a $\log M-\log T$ diagram implies
that the number of observed clusters $N(M,T)$ per unit age $dT$ in a
fixed $\log M$ range decreases inversely with age as $T^{-1}$ (Fall et
al. 2005). We refer to the slope of the number-age relation as $\chi$,
which would be $\chi=1$ in this case. Sometimes the decrease is a
little slower, e.g., as $T^{-0.7}$ ($\chi=0.7$; Whitmore, Chandar \&
Fall 2007; Mora et al. 2009), in which case the density of points on
the $\log M-\log T$ diagram increases slightly with $\log T$. In either
case, the decrease in cluster count with time is sometimes a power-law,
and this implies that clusters are dispersing or losing mass in a
regular fashion for a long time, usually for several hundred Myr.
Without cluster destruction, all clusters that ever formed would still
be present and the density of points on a $\log M-\log T$ diagram would
increase dramatically, in proportion to $T$, for a fixed $\log M$
interval.

Bastian et al. (2005a) studied the $\log M-\log T$ distribution for
clusters in M51 and concluded that there was, in fact, a significant
age gradient -- strong enough to be consistent with no cluster
destruction over time, only cluster fading. Hwang \& Lee (2010) studied
M51 again and found the same significant gradient with many more
clusters, leading to the same conclusion that cluster destruction is
minimal. Similarly, Peterson et al. (2009) show a $\log M-\log T$
diagram for Arp 284 and comment that they ``cannot rule out constant
cluster formation with no infant mortality,'' but they say this because
there are too few clusters to conclude either way.

Complicating this picture is the presence of local peaks in the number
of clusters along the $\log T$ axis, probably from short-term bursts.
M51 has such a peak corresponding to the time of its interaction with
NGC 5195 (Hwang \& Lee 2010), and the Antennae galaxy may have one too.
Bastian et al. (2009a) suggested that the uniform $\log T$ distribution
for constant $M$ in the Antennae reported by Fall et al. (2005) is
influenced by the on-going interaction, which caused the cluster
formation rate to increase globally by a factor of 10 between $\log
T=8.5$ and $\log T=7.5$, according to models by Mihos et al. (1993).
They also suggested that cluster distribution for shorter times is
dominated by cluster dispersal during gas removal. With these two
effects, there is no systematic $T^{-\chi}$ dispersal rate for clusters
in their model. Gieles \& Bastian (2008) also point out that the
maximum mass of clusters increases with $\log T$ for every galaxy with
adequate data, except for the Antennae. This increase implies that the
number of clusters in equal intervals of $\log T$ increases with $T$,
by the size-of-sample effect. Both of these observations counter the
interpretation that cluster numbers are about constant in equal $\log
T$ intervals.

If there is a power law fall-off of cluster count over time, then we
require $dM/dT\sim-\chi M/T$ for either evaporative-type (i.e., slow)
or disruptive-type (i.e., fast) cluster dispersal (e.g., Fall et al.
2009). This equation is contrary to expectations from standard
evaporation, where $dM/dT\sim$constant for the classical case (Spitzer
1987), and $dM/dT\propto -M^{0.38}$ for the Lamers et al. (2005) model.
The latter case may also be written
$dM/dT=-\left(M_0^{0.62}-0.62T\right)^{0.61}$ for initial mass $M_0$
after solving for $M(T)$ and differentiating with respect to $T$.
Neither of these cases has a cluster mass $\propto T^{-\chi}$ for a
decade or more in $T$. The observed distribution is also inconsistent
with cluster-cloud collisions at a constant mean free path, which would
predict an exponential decay in the number of clusters with age.

The usual interpretation for the time dependence in cluster counts
involves several distinct processes that occur at different phases in a
cluster's life. The youngest clusters become partially unbound when
their star-forming gas leaves (``infant mortality;'' Lada \& Lada
2003), depending on the efficiency of star formation in that gas and on
the relative rate of gas loss (e.g. Goodwin \& Bastian 2006; Baumgardt
\& Kroupa 2007). Clusters are also destroyed by collisions with dense
objects such as other clusters or molecular clouds (Gieles et al. 2006a
and references therein), they expand and lose stars after stellar mass
loss from winds and supernovae (Terlevich 1987), and they evaporate by
two-body relaxation (e.g., Baumgardt \& Makino 2003).  The puzzle is
that these four mechanisms generally have different time dependences,
and they occur at different times in the life of a cluster (see reviews
in Lamers \& Gieles 2006, Fall et al. 2009). There is no obvious reason
why they should combine to give an approximately power-law age
dependence inside a fixed mass interval.

The $\log {\rm M}-\log{\rm T}$ diagram may also be sectioned into fixed
intervals of $\log T$ to determine the shape of $N(M,T)$ for variations
in $M$. This shape is usually independent of $T$ for the observable
mass range, and approximately given by $N(M,T)\propto M^{-2}$ per unit
mass $dM$. This is the usual cluster mass function (Battinelli et al.
1994; Elmegreen \& Efremov 1997; Zhang \& Fall 1999; Hunter et al.
2003; de Grijs \& Anders 2006). The mass function has been observed in
many cluster populations and may follow from the distribution of mass
in dense cloud cores, which has about the same form (Reid \& Wilson
2005; Rathborne et al. 2006). Putting the $T$ and $M$ distributions
together implies that $d^2N(M,T)/dMdT \propto M^{-2}T^{-\chi}$ for
$\chi\sim0.7-1$ (Fall et al. 2009). If the mass distribution function
is steeper than $M^{-2}$ at high mass or has an exponential-like cutoff
at around $10^5$ to $10^6\;M_\odot$, then a Schechter mass function
might be more appropriate, giving $d^2N(M,T)/dMdT \propto
M^{-2}e^{-M/M_0}T^{-\chi}$ (Gieles et al. 2006bc; Waters et al. 2006;
Jordan et al. 2007; Bastian 2008; Gieles 2009; Larsen 2009).

The primary purpose of this paper is to study possible origins for the
time dependence of cluster counts. Our view is that cluster evolution
has to consider the large-scale star complex where most clusters form.
In a kpc-size complex, cluster dispersive forces gradually decrease as
the cluster drifts from its birthsite, and this automatically
introduces a power law time dependence for cluster dispersal.  We also
consider the effects of an upper mass cutoff, which seems to place
tight constraints on the dispersal mechanism. That is, if there is a
cutoff, then power-law $M-T$ relations are possible primarily in the
case where clusters are destroyed quickly, by cloud collisions, for
example. Finally, we investigate whether power law-type loss rates for
clusters might result from power law-like intensity profiles inside
clusters, in the sense that cluster mass is progressively lost from
view as the surface brightness fades below the limit of a survey.

In what follows, we first consider the importance of the cluster birth
environment, particularly the kpc-scale star complexes and their
hierarchical density structure (Sect. \ref{hier}). Then we model the
$\log M-\log T$ diagram in various ways, considering rapid disruption
as in the collisional model (Sect. \ref{instant}), slow erosion as in
the evaporation model (Sects. \ref{evap}, \ref{evap2}), combinations of
these two models (Sect. \ref{together}), cluster-cloud collisions
(Sect. \ref{coll}), cluster disruption by hierarchical disassembly
(Sect. \ref{disphier}), partial cluster disruption (Sect.
\ref{partial}), and apparent cluster mass loss by fading below the
surface brightness limit of a survey (Sects. \ref{false},
\ref{experiment}). In most cases, cluster loss as a power-law in time
can be reproduced with the right choice of parameters.  A summary of
the results is in Section \ref{disc}.

\section{The Importance of Hierarchical Birth Structure in the
Disruption Timing of Clusters}\label{hier}

Previous studies of cluster disruption have neglected the hierarchical
birth environment. Most star and cluster formation occurs in giant star
complexes that extend for an average of $\sim600$ pc in local galaxies
and last for $10^8$ years or more (Efremov 1995). These complexes and
their associated clouds (e.g., Grabelzky et al. 1987) are important for
cluster disruption because (1) the average cloud density exceeds the
tidal density from the background galactic potential by more than a
factor of 10 and therefore dominates the cluster evaporation rate, (2)
cloud pieces and other clusters are concentrated in a star complex and
dominate collisional disruption, and (3) hierarchically assembled loose
stellar groups can come apart in a hierarchical way. In all cases,
there is a power-law dependence in the physical structure of a young
cluster's environment, and this makes the cluster disruption rate vary
inversely with a power law of age as the cluster drifts through the
complex.

The mass dependence in $N(M,T)$ could also be the result of
hierarchical structure (Elmegreen \& Efremov 1997; Elmegreen 2008).
Hierarchical structure subdivides each cloud into $\xi$ smaller clouds,
preserving total mass $M$. The number $N$ of subclouds then increases
with level $L$ in the hierarchy as a power law $\xi^L$, and the mass of
each cloud decreases with each level as $M\xi^{-L}$. The product of the
mass and the number in each level is the total mass $M$, which is
constant. Because the levels are logarithmic in mass, we can write for
this total mass $MdN/d\log M=$constant, from which it follows that
$dN/dM\propto M^{-2}$ (Fleck 1996). We get the same result if clouds at
any level in the hierarchy are randomly selected with equal
probability.  This follows because at level $L$ there are $N(L)$
clouds, each with a mass $\propto1/N(L)$ on average. The probability of
selecting a certain mass is proportional to the number $N(L)$, and this
is $\propto1/M$. This probability is also proportional to the mass
function, which is therefore $\propto1/M$ for logarithmic intervals of
mass. A third model considers the packing density of objects in $D$
dimensions, where mass $M$ scales with size $R$ as $M\propto R^D$.  The
density is $n(k)dk\propto k^{D-1}dk$ for $k=1/R$ (Di Fazio 1986), and
since $M\propto k^{-D}$ and each mass corresponds to a definite $k$,
N(M)dM=n(k)dk, we get the mass function $N(M)\propto
k^{D-1}\left(dk/dM\right)=k^{2D}=M^{-2}$, independent of $D$. Numerical
experiments with fractal clouds demonstrate these results (St\"utzki et
al. 1998; Elmegreen 2002; Elmegreen et al. 2006).

Hierarchical structure is evident not only in the gas (Scalo 1985), but
also in the positions of young clusters (e.g., Zhang, Fall, \& Whitmore
2001; Scheepmaker et al. 2009), young stars (e.g., Gomez et al. 1993;
Elmegreen et al. 2003, 2006; Odekon 2006; Bastian et al. 2009b; Gieles,
Bastian, \& Ercolano 2008), and galactic HII regions (S\'anchez \&
Alfaro 2008), all of which have fractal structure and power-law
two-point correlation functions. Hierarchical structure continues even
inside embedded clusters (Gutermuth et al. 2005; Allen et al. 2007;
Schmeja, Kumar, \& Ferreira 2008; S\'anchez \& Alfaro 2009) and it is
present in the distribution of pre-stellar cores (Johnstone et al.
2000, 2001; Enoch et al. 2006; Young et al. 2006).  Cluster formation
is correlated in time also, such that clusters that are born closer to
each other are more similar in age (Efremov \& Elmegreen 1998; Fuente
Marcos \& de la Fuente Marcos 2009).  As a result, clusters form
grouped together (Piskunov et al. 2006; Bastian et al. 2007; de la
Fuente Marcos \& de la Fuente Marcos 2008) in star complexes (Efremov
1995) that span half a kiloparsec or more. Bound clusters appear to be
the densest part of the stellar hierarchy, where the local orbit time
is short enough to allow stellar mixing before gas dispersal (Elmegreen
2008; see reviews in Elmegreen 2009, 2010).

The ISM usually has an average gas density comparable to the tidal
limit, which is
\begin{equation}\rho_{\rm tidal}=-{{3\Omega
R}\over{2\pi G}} {{d \Omega}\over{dR}}\end{equation} for galactic
angular rotation rate $\Omega$ and galactocentric radius $R$. Locally,
$\rho_{tidal}\sim2.5m_H$ cm$^{-3}$ for Hydrogen mass $m_H$. Regions
with densities higher than $\rho_{tidal}$ are unstable in the absence
of pressure, so the excess gas can convert into stars if other
conditions are met. NGC 2366, for example, has an average ISM density
$\rho$ comparable to $\rho_{\rm tidal}$ for all radii, but in
star-forming regions, $\rho>\rho_{tidal}$ and outside of star-forming
regions, $\rho<\rho_{tidal}$ (Hunter et al. 2001).  Star complexes have
average gas densities above this limit, and the gas density gets
progressively higher closer to the cluster formation sites.  In the
inner part of the Milky Way for example, the gas density in each
$10^7\;M_\odot$ cloud complex is in the range of $5-10$ cm$^{-3}$
(Elmegreen \& Elmegreen 1987); in GMCs, the average density is
$\sim10^3m_H$ cm$^{-3}$ and in cluster-forming cores, it is
$\sim10^4-10^5m_H$ cm$^{-3}$.

The cluster formation environment is not uniform on any scale. As a
cluster drifts, it travels from an initially high density region where
the tidal forces and collision rates are large, to a low-density region
where the tidal forces and collision rates are small.  For kpc-size
star complexes, this migration can take 100 Myr or more. Cloud
complexes have density profiles between $\rho\propto S^{-2}$
(isothermal), and $\rho\propto S^{-1}$ (Larson's law), for distance
$S$, so the tidal density and collision rate vary with distance in this
way.

The tidal force gradient determines the cluster tidal radius. Because
the cluster evaporation rate is inversely proportional to the crossing
time inside this radius, the evaporation rate scales with the square
root of the environmental density.  Note that stellar orbits inside a
cluster have much shorter periods than any of the other time scales we
are discussing, so the cluster should adjust internally as it moves
through different tidal fields. For a cluster younger than several
times $10^8$ years, the environmental density is dominated by the cloud
complex in which the cluster formed, so the evaporation rate could
decrease as $S^{-1}$ or $S^{-1/2}$, given the two density profiles
above. If the cluster drift speed is constant, then $S\propto T$ and
the instantaneous evaporation rate is $\propto T^{-1}$ or $T^{-1/2}$.
At the same time, the cluster collision rate decreases with time as the
cluster drifts from its birth site, in direct proportion to the clump
density, which is $\propto T^{-2}$ or $T^{-1}$ for these two radial
profiles, respectively.

An evaporation rate $dM/dT$ that varies with cluster mass and inversely
with cluster age has the property that the resulting cluster $M-T$
distribution is uniform over $\log T$ in fixed intervals of cluster
mass.  Similarly, a cluster collisional disruption rate that scales
inversely with $T$ produces the same uniform $\log T$ distribution on
such a plot. Any combination of these two cluster disruption mechanisms
also has this form.  We demonstrate these distributions in Section
\ref{models}.

Such a model for evaporation is not standard, however. We need
$dM/dT\sim-\chi M/T$ to get the proposed near-uniformity on a $\log
M-\log T$ distribution and standard evaporation has
$dM/dT\sim$constant, i.e., without either the mass dependence or the
age dependence. Cloudy structure with a background density gradient, if
appropriate, would contribute only the $\sim1/T$ dependence to this
differential; it would not introduce the required mass dependence.
Cloud disruption by a $1/T$ collision rate could solve the problem
because there is automatically a mass dependence there: disruption
removes the whole mass on the collision timescale, giving the required
change in mass, $\Delta M=-M$ per unit collision time.  Thus disruption
of clusters by cloudy debris in a star complex may be favored over
evaporation. However, this does not mean that a different kind of slow
cluster disruption is not happening also. One can imagine a type of
cluster harassment by repetitive tidal forces from cloudy debris and
other clusters that energize the outermost cluster stars and lead to a
slow but global loss from the cluster. For such a global process,
$\Delta M=-M$ as required, and with an ever-decreasing harassment
frequency, proportional to the local density of colliding cloud debris,
we might get $dM/dT\propto -\chi M/T$.

A second aspect of hierarchical cluster formation is that some cluster
disruption could be hierarchical too, with large centers drifting apart
from each other if they are not mutually bound by gravity and small
sub-centers inside each one drifting apart from each other on a
different timescale. In surveys of clusters or star-forming regions
with poor spatial resolution, such as surveys in other galaxies, what
is sometimes called a cluster may be only a collection of smaller
clusters or unbound stars. Expansion of these regions then changes the
unresolved object that is identified as a cluster, lowering its mass or
making two small clusters instead of one large cluster. Thus another
disruption mechanism for clusters is the hierarchical disassociation of
substructures. Such disassociation should be accompanied by stellar
dispersal into the field (e.g., Gieles et al. 2008; Bastian, et al.
2009b).  Hierarchical dispersal gives a uniform distribution on a $\log
M - \log T$ plot under the conditions discussed in Section
\ref{disphier}.

The importance of the cluster formation environment over the average
galactic environment for collisional impact disruption can be seen from
the theory in Gieles et al. (2006a). They use a cluster mass loss rate
of $dM/dt=-M/t_{dis}$ where, for GMC collisions,
\begin{equation}
t_{dis}=37{M^{0.61}\over{\Sigma_n\rho_n}}\; {\rm Myr}.\label{dis}
\end{equation}
Here $\Sigma_n$ is the average column density of a colliding cloud in
$M_\odot\;{\rm pc}^{-2}$, $\rho_n$ is the average density of gas in the
neighborhood, in $M_\odot\;{\rm pc}^{-3}$, and $M$ is the cluster mass.
They choose $\Sigma_n=170M_\odot\;{\rm pc}^{-2}$ for typical GMCs and
$\rho_n=0.03M_\odot\;{\rm pc}^{-3}$ for the ambient ISM. They were
thinking of clusters moving in the average ISM, colliding with GMCs now
and then. However, the mass of the colliding cloud does not enter
$t_{dis}$, and the collision partners could be pieces of a cloud
complex as well. As shown by Larson (1981) and Solomon et al. (1987),
most molecular clouds have $\Sigma_n=170M_\odot\;{\rm pc}^{-2}$
regardless of their mass, whereas standard diffuse clouds with 1
magnitude of extinction have $\Sigma_n=10M_\odot\;{\rm pc}^{-2}$. We
therefore expect that the cloud pieces a cluster meets during its first
several hundred Myr will have $\Sigma_n$ starting near
$\sim170M_\odot\;{\rm pc}^{-2}$ and ending near $\sim10M_\odot\;{\rm
pc}^{-2}$. At the same time, the density $\rho_n$ in the cluster's
neighborhood starts with the high value of its formation site,
$\rho_n\sim60M_\odot\;{\rm pc}^{-3}$ ($=10^3$ H$_2$ cm$^{-3}$) and
drops by two or three orders of magnitude as the cluster drifts. These
numbers imply that $\Sigma_n\rho_n$ might start near
$10^4M_\odot^2\;{\rm pc}^{-5}$ when the cluster first emerges from its
forming cloud core, and steadily drop to a final value of
$\sim10M_\odot^2\;{\rm pc}^{-5}$ or lower as the cluster leaves the
complex. Eventually the cluster drifts into the field where
$\Sigma_n\rho_n\sim5.1M_\odot^2\;{\rm pc}^{-5}$, as assumed by Gieles
et al. (2006a).

Cluster disruption by collisions with nearby GMC clumps is similar to
cluster disruption by rapid gas loss. Both involve sudden changes in
the gravitational potential near the cluster. Thus the transition from
``infant mortality'' in the sense of wind- and ionization-driven gas
loss to ``adolescent mortality'' in the sense of cluster- GMC clump
collisions, could be rather smooth. Whether the disruption is slow from
repetitive collisions or rapid from a single strong collision, the
$\log M-\log T$ diagram will be the same as long as both produce a
time-average mass loss like $dM/dT\sim-\chi M/T$. Most likely, the
disruption depends on the actual sequence of tidal forces from the
particular cloud clumps that a cluster encounters. Some clusters could
disperse slowly following many weak collisions, while others could
disperse little at first and then suddenly come apart following a
single strong collision. There could also be a gradual expansion of
clusters that accompanies steady disruption (e.g., Bastian et al. 2008;
Wilkinson et al. 2003). All of these situations are modeled in the
following sections.

\section{Models of the $\log M - \log T$ Plot for Clusters}\label{models}

Several types of models that generate an approximately constant density
of clusters on a $\log M - \log T$ plot are discussed here. All of the
models generate clusters with a randomly chosen mass as time passes,
with one new cluster per time step in a fixed time interval $dt$. Thus
the cluster formation rate is one new cluster per interval of time
$dt$. The mass distribution function of these newly generated clusters
is the standard power law, $N(M)dM\propto M^{-2}dM$, from a minimum
cluster mass of $10\;M_\odot$ to a maximum cluster mass of
$10^6\;M_\odot$ ($10^8\;M_\odot$ for one model described below). Models
with Schechter mass functions are shown for comparison in three cases.
Clusters are destroyed by various means after their age is 1 Myr. The
distribution of cluster mass and age is then shown on a $\log M-\log T$
plot after a sufficiently large number of steps (e.g., 50,000). We do
not consider fading limitations but assume that all of the clusters
plotted are above the fading limit. The fading limit can be avoided
similarly in real observations by considering only clusters more
massive than the minimum detectible mass at the oldest age of interest
(e.g., Melena et al. 2009). Fading is considered in more detail in
Section \ref{false}.

Analytical treatments of $N(M,T)$ for instantaneous disruption and
smooth cluster mass loss were presented by Fall et al. (2009).  They
did not generate stochastic $\log M-\log T$ diagrams nor consider the
cases discussed in Sections \ref{together}--\ref{false} below.

\subsection{Instantaneous Cluster Disruption with a Rate Proportional to
the Inverse of the Cluster Age}\label{instant}

The first model destroys clusters instantly with a probability $P$,
which is proportional to the disruption rate per cluster. For each time
step, we loop through all of the existing clusters, find the current
age of each one (the age is the difference between the current time and
the cluster formation time) and then assign a probability of its
disruption in that time step equal to some constant multiplied by the
time interval $dt$, and divided by the age $T$,
\begin{equation}
dP_{\rm dest}=\chi_1 dt/T.\label{eq:1}\end{equation} Thus the
instantaneous disruption rate for a cluster of age $T$ is $\chi_1/T$.
We determine if a particular cluster is actually destroyed by
generating a random number uniformly distributed between 0 and 1 for
that cluster and comparing it to $dP_{\rm dest}$. If the random number
is less than $dP_{\rm dest}$, then that cluster is destroyed, which
means it is removed from the list of all clusters. This exercise of
generating a $dP_{\rm dest}$ and a random number is done for every
existing cluster at each time step, and after that time step, all of
the destroyed clusters have been removed from the list of existing
clusters.

The constant $\chi_1$ affects the gradient in the density of points on
a $\log M - \log T$ plot. When $\chi_1=1$, this density of points is
constant in the $T$ direction. If $\chi_1<1$, then the disruption rate
is low and there is an overabundance of old clusters compared to young
clusters. If $\chi_1>1$, then the disruption rate is high and there is
an overabundance of young clusters compared to old clusters.

\begin{figure}
\epsscale{1.0} \plotone{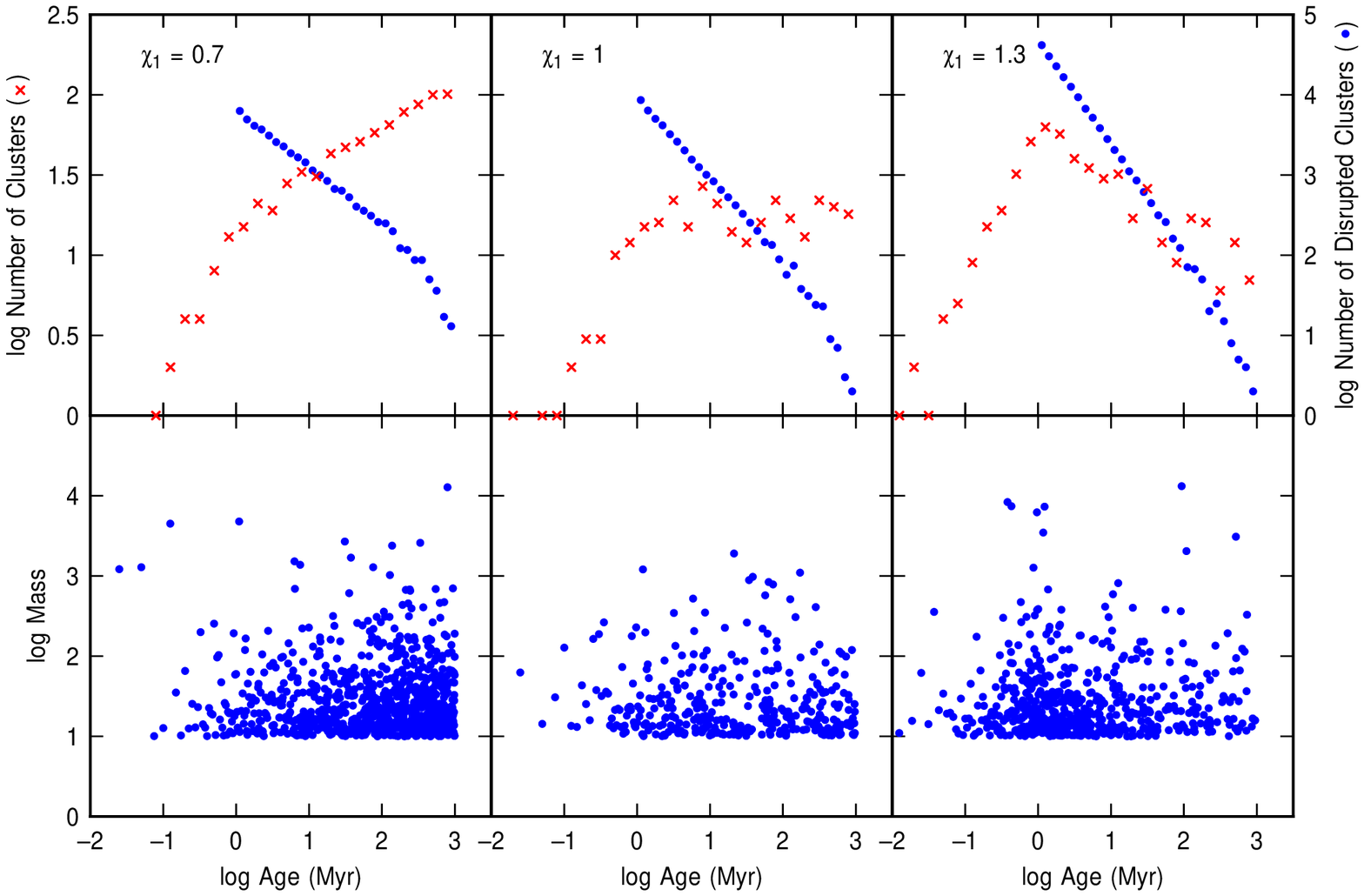} \caption{(Bottom) Cluster age and mass
distributions for a model with sudden and complete cluster disruption
and a probability per timestep given by equation (\ref{eq:1}). (Top,
red crosses) The number of clusters in the mass range from $\log
M/M_\odot=1$ to 2 in equal intervals of $\log T$. (Top, dots) The
number of disrupted clusters as a function of their age at disruption.
The age distribution is sensitive to $\chi_1$. Each cluster has a track
on this plot that is a straight horizontal line with a beginning far to
the left and an end at the age of disruption. \label{clusdest}}
\end{figure}

The lower panels of Figure \ref{clusdest} show the distributions of
clusters on $\log M - \log T$ diagrams for this instantaneous
disruption model. On the left, $\chi_1=0.7$, in the middle $\chi_1=1$,
and on the right $\chi_1=1.3$.  The three top panels show the density
of points in the lower panels between $\log M=1$ and $\log M=2$, as red
crosses (using the left-hand axes), measured in bins of equal $\log T$
intervals. The density of points is constant when $\chi_1=1$. The slope
of the trend shown by the red crosses is $1-\chi_1$.  This can be seen
from the differential equation that is equivalent to the random
sampling model given in equation \ref{eq:1}, namely, $dM/dT=-\chi_1
M/T$, which has the solution $M(T)=M_0(T/T_0)^{-\chi_1}$ for starting
mass $M_0$ and time $T_0$. In a logarithmic box of size $d\log M \times
d\log T$ around $M$ and $T$, the clusters were born with a mass
$M_0=M\left(T/T_0\right)^{\chi_1}$ in the interval $d\log M_0=d\log M$.
The rate at which they were born is inversely proportional to $M_0$ for
the initial cluster mass function $dn(M_0)/d\log M_0\propto M_0^{-1}$,
and the number of them remaining at time $T$ in the interval $d\log T$
is the rate at $T_0$ times the time interval represented by $d\log T$,
which is the number in $d\log T_0$ times $T/T_0$. Thus the number in
the interval $d\log M\times d\log T$ is proportional to
$\left(T/T_0\right)^{-\chi_1}\times\left(T/T_0\right)=
\left(T/T_0\right)^{1-\chi_1}$.

The absolute value of the density in the $\log M - \log T$ plot is
determined by the value of $dt$: lower $dt$ corresponds to a higher
formation and disruption rate, and more points in the plot.  For the
panels from left to right, $dt=0.025$, 0.025, and 0.00625. The blue
dots in the top panels show the age distribution functions of all the
destroyed clusters using the right-hand axes; these are the clusters
that were removed from the running list during the simulation because
their random number was less than their $dP_{\rm dest}$ value at the
time of their disruption.  The slope of the age distribution for
destroyed clusters gets steeper as $\chi_1$ increases. This is because
the disruption rate is then higher, more clusters are destroyed, and so
the number of clusters falls off more rapidly with age. For the same
reason, $dt$ has to be smaller for larger $\chi_1$ to have about the
same number of clusters today.

The maximum mass of a cluster in each bin of $\log T$ tends to follow
the number of clusters in the bin by the size-of-sample effect. When
the number is constant, as indicated by a horizontal distribution of
red crosses in the top panel, then the upper limit of the blue dots in
the lower panel is constant also. Similarly, when one increases with
$T$, the other increases too, in direct proportion. We have observed
this effect in dwarf irregular galaxies (Melena et al. 2009).

The linear relationship between peak cluster mass and cluster number is
a consequence of the cluster mass function $dN/dM\propto M^{-2}$. This
may be seen by setting $\int_{M_{max}}^\infty N(M)dM=1$, which means
that there is one cluster with a maximum mass of $M_{\rm max}$ or
larger. In this case, the total number of clusters above a certain mass
$M_{\rm min}$ equals $\int_{M_{\rm min}}^{M_{\rm max}}N(M)dM=M_{\rm
max}/M_{\rm min}$. That is, the number is directly proportional to
$M_{\rm max}$ for fixed $M_{\rm min}$.

We observed a different trend in the LMC (Hunter et al. 2003), where
the maximum mass of a cluster increased with age even though the number
of clusters in bins of equal $\log T$ for a given range of $\log M$ was
about constant. Gieles \& Bastian (2008) also observed this seemingly
contradictory effect in several other galaxies. In Hunter et al.
(2003), we used the rate of increase in maximum mass to infer the
cluster mass function, suggesting that these largest clusters were not
as likely to be destroyed as lower mass clusters, and were therefore
still an indication of the birth mass function. Gieles \& Bastian used
the maximum mass trend to infer that the $dN/dT\propto 1/T$ relation
does not apply in some cases.

We return to this point in Sections \ref{evap2} and \ref{coll} where
models that get both a constant density over $T$ in a $\log M-\log T$
plot and a linearly increasing $M_{\rm max}(T)$ are shown.  What breaks
down is the $dN/dM\propto M^{-2}$ mass function for all ages. This
breakdown is a consequence of an assumed cluster mass dependence in the
disruption time. The models in the present section have a disruption
time independent of cluster mass so $dN/dM\propto M^{-2}$ is preserved
and the breakdown is not seen.

\begin{figure}
\epsscale{1.0} \plotone{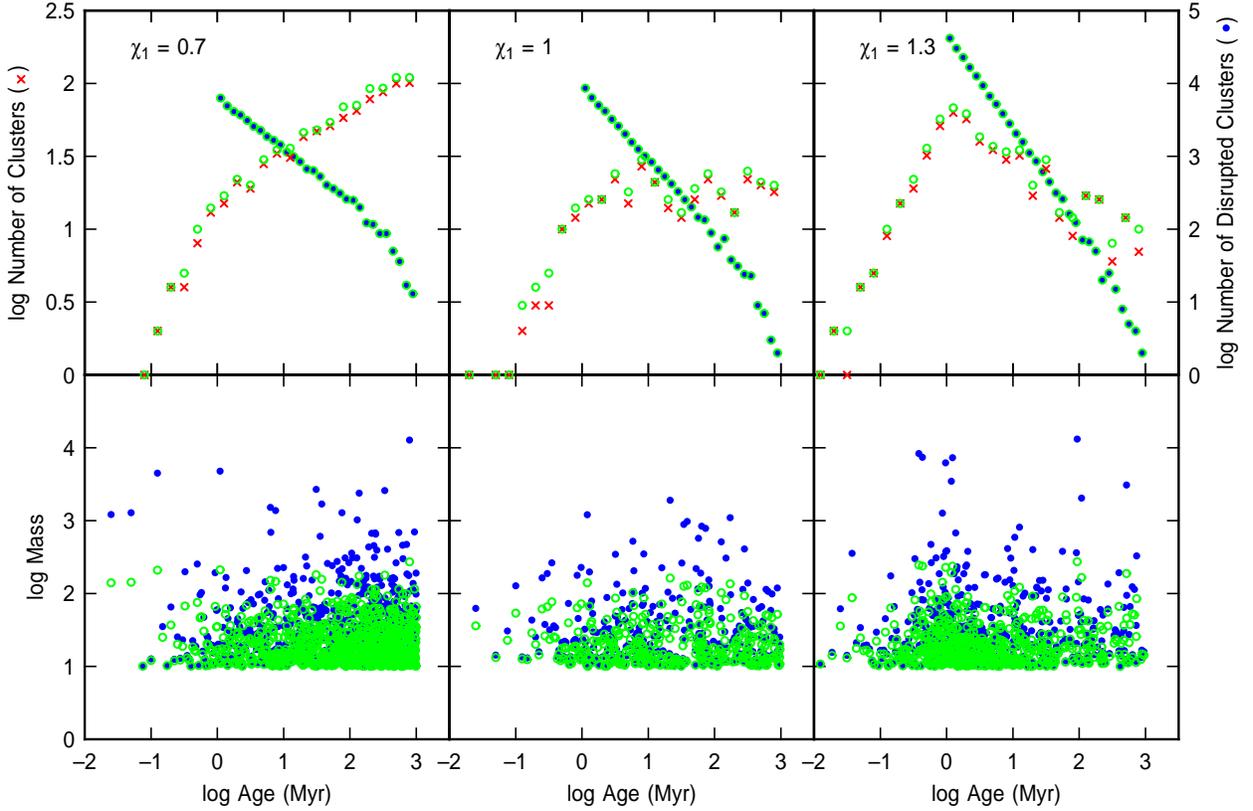} \caption{(Bottom:) Cluster age and mass
distributions repeated from Figure 1 (blue dots), for which a power law
cluster mass function was assumed, compared to the age and mass
distributions with a Schechter cluster mass function (green circles).
The cutoff mass is $M_0=100\;M_\odot$, low enough to observe on this
diagram with relatively few clusters. The upper cutoff has little
effect on the distributions aside from lowering the masses to near and
below the cutoff. (Top, red crosses and blue dots:) The number of
clusters in the mass range from $\log M/M_\odot=1$ to 2 in equal
intervals of $\log T$, and the number of disrupted clusters, again as
in Figure 1.  The green circles are for the Schechter function. The
number of disrupted clusters (power law) is exactly the same in the two
cases, so the green circles overlay the blue dots in the upper panels.
The age distribution does not change significantly when there is a
cluster mass cutoff. \label{clusdest_schechter}}
\end{figure}

Figure \ref{clusdest_schechter} shows the effect of an upper mass
cutoff in the Schechter (1976) mass distribution function using the
instantaneous destruction model.  The blue dots in the bottom panels
repeat the distributions in Figure \ref{clusdest} and the red crosses
in the top panels repeat the density profiles in that figure. The green
circles in all panels are for a Schechter cluster mass function instead
of a power law. The Schechter function is $dN/dM\propto
M^{-2}e^{-M/M_0}$ for upper cutoff mass $M_0=100\;M_\odot$. The
observed cutoffs for real galaxies are around $10^5\;M_\odot$ or
$10^6M_\odot$ (Gieles et al. 2006c; Gieles 2009; Bastian 2008; Larsen
2009), but our cluster samples are not large enough to get into this
rare cluster range, so we pick a much lower $M_0$ to study the effect.

The figure shows that an upper mass cutoff has very little effect on
the distribution of points in the $\log M-\log T$ diagram for a rapid
dispersal model with $T^{-\chi}$ dispersal probability. This is because
the trajectory of points on this diagram is purely horizontal: clusters
preserve their mass as they age and then disappear suddenly.  The main
effect is that the cluster masses in the lower panels decrease a
little, and the density of clusters per unit $\log T$ interval
increases a little for the $\log M=1$ to 2 mass range. The increase
occurs because clusters more massive than $10^2\;M_\odot$ for the
power-law case are now in the mass range used to calculate the density.
The mass distribution of dispersed clusters in the top panel is exactly
the same as in Figure \ref{clusdest} because the same random numbers
are used in the two cases for both the initial mass sequence and the
dispersal probability.

\subsection{Slow Cluster Mass Loss with a Rate Inversely
Proportional to Cluster Age}\label{evap}

\begin{figure}
\epsscale{1.0} \plotone{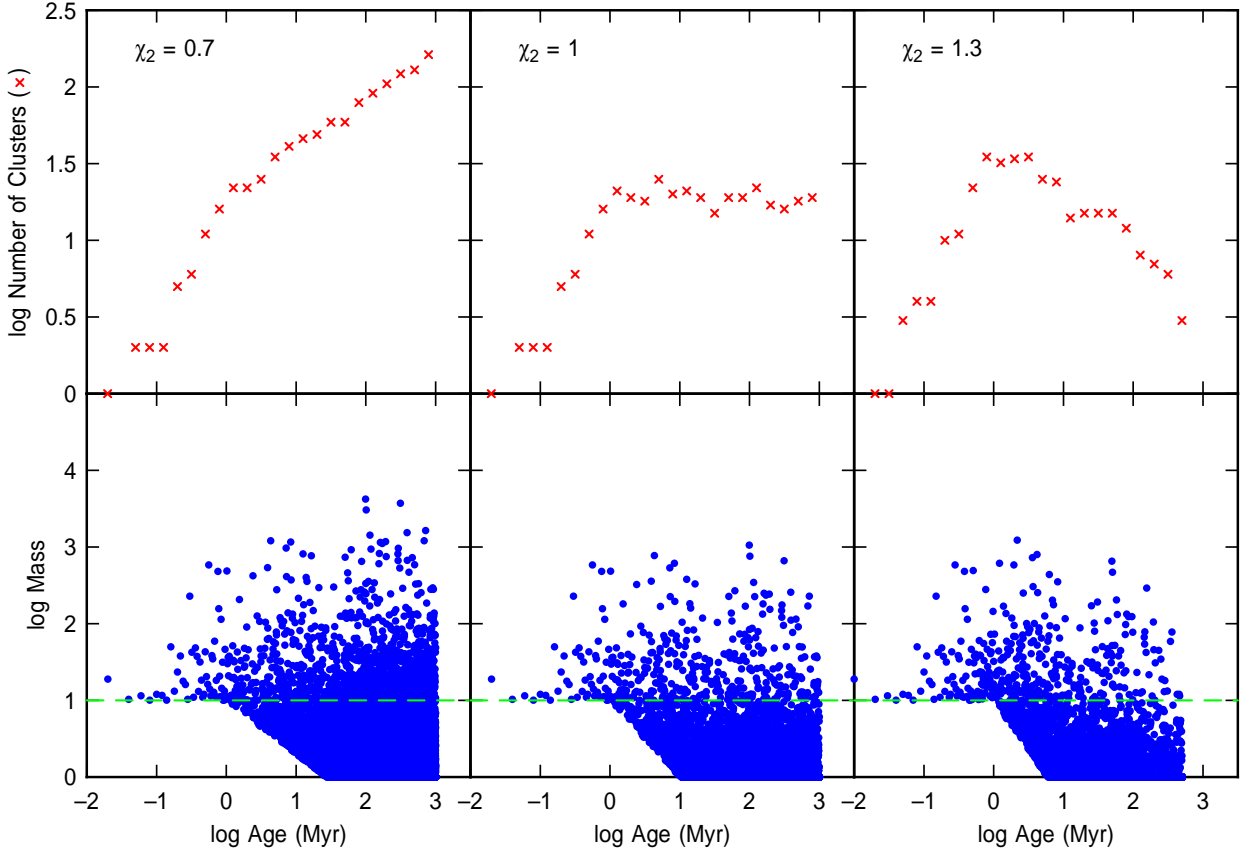} \caption{(Bottom) Cluster age and mass
distributions for a model with continuous cluster mass loss given by
equation (\ref{eq:2}). (Top) The number of clusters in the mass range
from $\log M/M_\odot=1$ to 2 in equal intervals of $\log T$. The age
distribution is sensitive to $\chi_2$. Each cluster has a track on this
plot that is a straight line with a slope $-\chi_2$, parallel to the
lower edge of the distribution at $\log M<1$ and $\log T>0$. All
clusters have an initial mass larger than $\log M=1$.
\label{clusdest2}}
\end{figure}

The second model erodes each cluster by a small amount at each time
step, as may be the case for cluster mass loss from stellar evolution,
cluster evaporation, and cluster harassment. Clusters are randomly
chosen from an $M^{-2}$ mass function at a rate of one per time
interval $dt$, like before. For each time step, we loop over all
existing clusters and decrease their mass with a mass loss rate
\begin{equation}
dM/dt = -\chi_2M/T .\label{eq:2}\end{equation} Clusters never disappear
in this case, they just get lower in mass and drop off the bottom of
the $\log M - \log T$ diagram. From an observational point of view, the
bottom of the $\log M-\log T$ diagram is the maximum mass that is
detectible with high confidence among all of the possible ages that are
considered. The constant $\chi_2$ acts like $\chi_1$ in the previous
example. When $\chi_2$ is less (greater) than 1, the disruption rate is
low (high) and there are too many (few) old clusters compared to young
clusters. Figure \ref{clusdest2} shows the results. The dashed green
line at $\log M=1$ indicates the lower mass cutoff for the
observational selection of clusters (and the selection of initial
clusters in our models). In Figure \ref{clusdest}, clusters never got
less massive than this because they either stayed the same or were
destroyed all at once. In Figure \ref{clusdest2}, all clusters lose
mass continuously and eventually drop below this initial minimum. The
slope of the lower mass border is parallel to the slopes of all the
cluster trajectories, which are downward and to the right at an angle
from the horizontal equal to $\arctan\chi$. For the 3 cases in the
figure, these angles are $35^\circ$, $45^\circ$, and $52^\circ$. The
values of $dt$ in the three cases are $0.02$, $0.02$, and $0.01$,
respectively.

The maximum mass in Figure \ref{clusdest2} increases or decreases with
the total number of clusters, as in Figure \ref{clusdest}, by the size
of sample effect. This is the case when the disruption time is
independent of the cluster mass.

\begin{figure}
\epsscale{1.0} \plotone{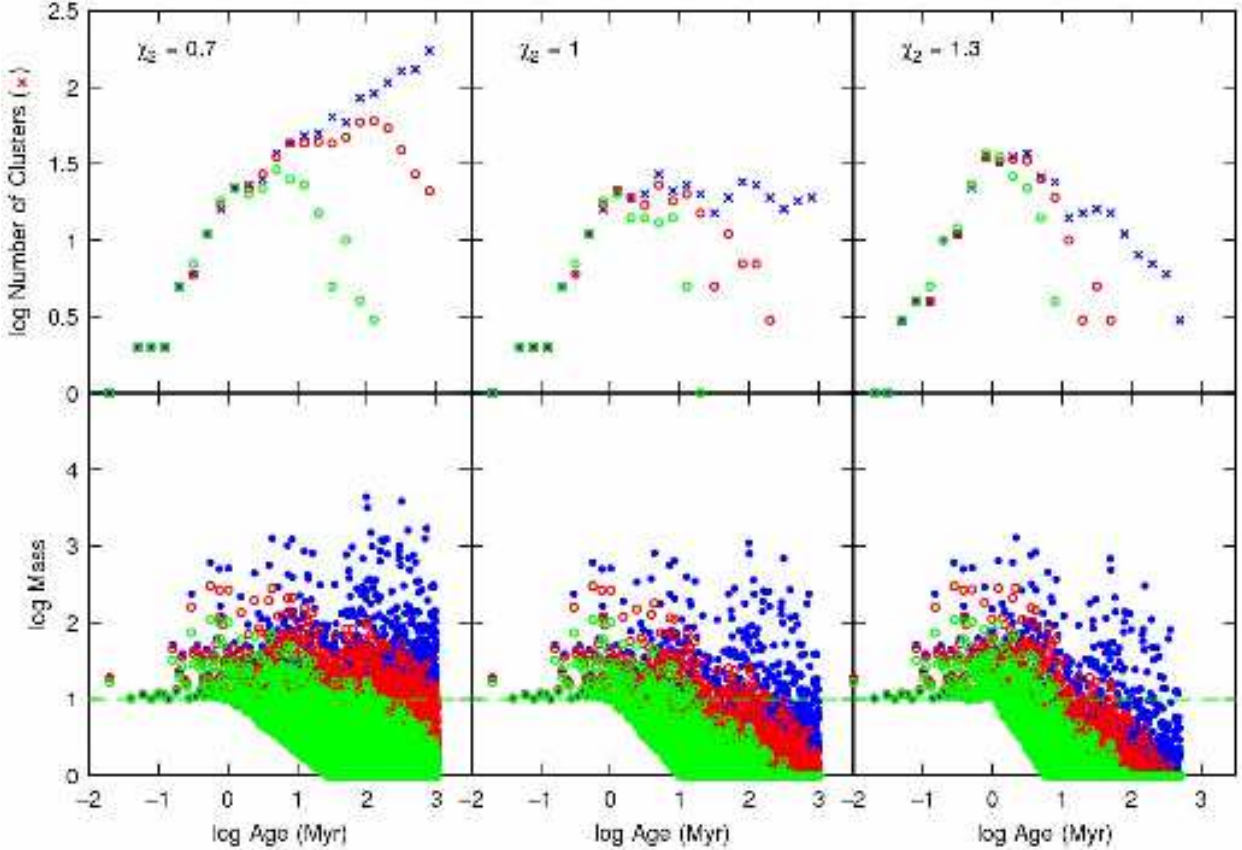} \caption{(Bottom) Cluster age and mass
distributions for the same model as in Figure 3 but with a Schechter
mass function. The blue points in the bottom and the blue crosses in
the top are the same as the blue points and red crosses in Figure 3.
The red circles have a cutoff mass $M_0=10^3\;M_\odot$ and the green
circles have a cutoff mass $M_0=10^2\;M_\odot$.  In this figure, red
circles overlay blue dots and green circles overlay both; thus one
should imagine there are blue dots beneath the red and green circles,
and red circles beneath the green circles, i.e., all symbols go down to
the same lower mass limit at each age. The mass cutoff significantly
affects the distribution of points on this diagram, suggesting that a
cutoff and the model with slow power-law disruption are mutually
inconsistent for some cluster populations. (Image degraded for arXiv.)
\label{clusdest2_schechter}}
\end{figure}

The effect of an upper cutoff mass is shown in Figure
\ref{clusdest2_schechter}.  The blue dots in the bottom panels and the
blue crosses in the top panels are the same as the blue dots and red
crosses in Figure \ref{clusdest2}, but now superposed on these are red
and green circles showing cluster populations with cutoff masses in a
Schechter function of $M_0=10^3$ and $10^2\;M_\odot$ respectively.  For
slow cluster dispersal where $M\propto T^{-\chi}$, the cutoff mass
affects the distribution of clusters on a $\log M - \log T$ diagram.
This is because the trajectory of points on this diagram is sloping
downward and to the right, so the clusters at old age and intermediate
mass were formerly at higher masses. For the $\chi=1$ case, each
position $(\log M,\log T)$ on the diagram traces back to a birth
position $(\log M + \log T/T_0, \log T_0)$. If $\log M+\log T/T_0$ is
higher than the cutoff mass, then that position will be nearly empty on
the diagram. For general $\chi$, the birth position is $\log M+\chi
\log T/T_0$.

Distributions of clusters like the red and green circles in Figure
\ref{clusdest2_schechter} are not generally observed.  The observations
of fairly uniform $\log M-\log T$ diagrams suggest that if there is an
upper mass cutoff in the range of $10^5-10^6\;M_\odot$, then cluster
disruption has to be rapid, as in the cloud collision model.
Alternatively, if each cluster disrupts steadily with a $T^{-\chi}$
mass loss rate, then the upper mass cutoff at age $T_0$ has to be
higher than $\log M_{max} + \chi\log T_{max}/T_0$ for the most massive,
$M_{max}$, and oldest, $T_{max}$, clusters that are observed.

\subsection{Slow Cluster Mass Loss in the Lamers, Anders, \& de Grijs
Evaporation Model}\label{evap2}

Figure \ref{clusdest2_lamersfit} shows another study of cluster slow
mass loss. The left two panels repeat the $\chi_2=1$ case from Figure
\ref{clusdest2}, now with 4 curves drawn on the bottom panel showing
the decay tracks of 4 clusters. The decay is at a 45 degree angle in
this plot, which is why the $dN/d\log M\propto 1/M$ mass distribution
for clusters converts exactly into a $dN/d\log T =$ constant age
distribution.

The middle two panels of Figure \ref{clusdest2_lamersfit} show the
cluster distribution in the model of evaporation discussed by Lamers,
Anders, \& de Grijs (2006), in which there is a first phase of mass
loss from stellar evolution followed by a second phase of mass loss
from evaporation. We use the same fiducial disruption time here as in
that model, $t_0=21.8$ Myr, for these middle panels. Then
$dM/dt=-M/t_{dis}$ from the evaporation part of the disruption, where
$t_{dis}=t_0M^{0.62}$. The curves in the bottom panel follow individual
clusters again. They show the general properties of cluster
evaporation: the individual cluster masses decrease relatively slowly
at first in a log-log plot, and then at an ever increasing rate as the
cluster mass approaches zero. The same is true for the standard model
of cluster evaporation, which has a cluster mass decrease linearly with
time (Spitzer 1987). Such cluster mass loss does not give a
horizontally-constant distribution of clusters on a $\log M-\log T$
plot: the density increases rapidly for higher age $T$, as shown in the
top panel by the rising density of clusters per unit $\log T$.

The right two panels in Figure \ref{clusdest2_lamersfit} show the
Lamers et al. (2006) model again but with $t_{dis}=0.1t_0TM^{0.62}$ for
cluster age $T$. That is, $t_{dis}$ starts much smaller than in the
Lamers et al. model and increases with age, as in the left-hand panel.
Such an age dependence is not physically realistic for cluster
evaporation in a uniform tidal field, but it may apply in a varying
density field, as discussed in Section \ref{hier}. We call this the
modified Lamers model. It has the nice result that the distribution of
points on a $\log M-\log T$ plot is more uniform than without the age
dependence because the initial decline of cluster mass is fast,
somewhat like the 45 degree angle in the left-hand plot. By the time
the decline in mass becomes much steeper than this, the clusters are
below the observational limit (the green dashed line) and the resulting
non-uniformity in the age distribution cannot be observed. Note that
this steepening of the mass decay curves is intrinsic to the
conventional Lamers et al. and Spitzer models because $t_{dis}$
decreases as the mass decreases in both cases. However, the modified
Lamer et al. model gives a more uniform distribution on the $\log
M-\log T$ plot because the $M^{0.38}$ mass dependence in $dM/dt$ is
partially compensated by a $T$ dependence in $t_{dis}$. A similarly
modified Spitzer model would be less uniform on a $\log M - \log T$
plot than the modified Lamers model because there is no mass dependence
in $dM/dT$ for the Spitzer model.

We note that in both the Lamers et al. model and the modified Lamers et
al. model, the maximum mass of a cluster, $M_{\rm max}$, increases with
age, even in the right-hand plot where the density of points in a fixed
interval of $\log M$ is about constant over $T$. The reason for this is
that with $t_{dis}\propto M^{0.62}$ or any other positive power of $M$,
high mass clusters get destroyed proportionally slower than low mass
clusters, so the high mass clusters stay around longer and contribute
to the rising $M_{\rm max}$, even as the number of low mass clusters
drops rapidly. The implication is that the cluster mass function must
get flatter over time if $M_{\rm max}$ increases and $dN/dT\propto
1/T$. Hwang \& Lee (2010) observe a flattening of the cluster mass
function with time in a case like this. The mass function could even
develop a low-mass turnover with time as it flattens in the middle-mass
range. We return to this point in Section \ref{coll}.

Figure \ref{clusdest2_lamersfit4} shows the effect of an upper mass
cutoff in the Lamers, Anders, \& de Grijs (2006) model for stellar
evolution and cluster evaporation. The left two panels repeat the
results from the middle panel of Figure \ref{clusdest2_lamersfit}, for
ease of comparison. The middle and right pairs of panels in Figure
\ref{clusdest2_lamersfit4} are for upper mass cutoffs
$M_0=10^3\;M_\odot$ and $10^2\;M_\odot$, respectively, in the mass
function $dN/dM\propto M^{-2}e^{-M/M_0}$.  The figure shows that an
upper mass cutoff has little effect on the distribution of points in
the $\log M-\log T$ diagram because the movement of clusters in this
diagram is mostly from left to right until the mass gets quite low.
Thus clusters at high $\log T$ are not generally from those former
clusters that had high $M$ when they were born.

\subsection{Slow Cluster Mass Loss and Rapid Disruption
Together}\label{together}

Another model related to the first two has all of the clusters lose
mass slowly according to equation (\ref{eq:2}), while some of the
clusters are destroyed rapidly with a rate given by equation
(\ref{eq:1}). In this case, the effective slope of the $M(T)$ relation
is $\chi_1+\chi_2$. This implies that if instantaneous disruption from
strong cloud or cluster collisions is too slow to give an approximately
constant density alone ($\chi_1<1$), then slow mass loss from cluster
harassment can make up for the difference by decreasing the mass of
each one continuously.

Figure \ref{clusdest3} shows three cases. On the left is $\chi_1=0.3$
and $\chi_2=0.4$, giving a total $\chi_1+\chi_2=0.7$.  In the middle is
$\chi_1=0.6$ and $\chi_2=0.4$, whose sum is 1. On the right is
$\chi_1=0.3$ and $\chi_2=0.7$, whose sum is also 1. The time interval
is $dt=0.02$ in all cases.  We do not show a case with
$\chi_1+\chi_2>1$ because the result looks like that in the right-hand
panels of Figures \ref{clusdest} and \ref{clusdest2}. The density of
points on the $\log M - \log T$ diagram is constant in the second two
cases, which both have $\chi_1+\chi_2=1$. The age distribution of
destroyed clusters (blue dots, top panels, and right-hand axes) differs
in those two cases, though. Also, the rate of decrease of minimum
cluster mass with age below the initial lower limit (in the bottom
panels of Figure \ref{clusdest3}), is proportional to $\chi_2$.

In Figure \ref{clusdest3}, the maximum cluster mass follows the same
trend as the number of clusters because the disruption time is
independent of cluster mass.

\subsection{Cluster Disruption by Cloud Collisions}\label{coll}

The Gieles et al. (2006a) model of cluster-GMC collisions suggests that
the disruption rate of clusters is given by the equation
$dM/dt=-M/t_{dis}$ with $t_{dis}$ written above in equation
(\ref{dis}). This model does not have the general form required for a
power-law $M(T)$ relation, which is $dM/dt=-\chi M/T$. We can modify
the Gieles et al. model to have approximately this form, however, by
allowing the density of collision partners to decrease with time as the
cluster moves through the star and cloud complex in which it was born
(Sect. \ref{hier}). Using the numbers given in Section \ref{instant}
for the product $\Sigma_n\rho_n$, namely, a variation from
$10^4\;M^2_\odot\;{\rm pc}^{-5}$ to $10\;M^2_\odot\;{\rm pc}^{-5}$
between, say, 1 Myr and $10^3$ Myr, and assuming that this product
decreases inversely with time because of the power law structure in the
initial kpc-size cloud, we assign
$\Sigma_n\rho_n=10^4/T\;M^2_\odot\;{\rm pc}^{-5}{\rm Myr}^{-1}$ for $T$
in units of Myr. Then
\begin{equation}
t_{dis}=0.0037\;T M^{0.62} \;{\rm Myr}.\label{dis2}
\end{equation}

Figure \ref{clusdest7} shows the results from this case. The cluster
formation rate has to be extremely high to compensate for the high
disruption rate given by equation (\ref{dis2}); we take $dt=0.0001$ in
the left two panels, and $dt=0.0005$ (5 times lower formation rate) in
the middle two panels. In both cases the maximum cluster mass is
$10^8\;M_\odot$ because the cluster disruption rate is very high (an
upper cutoff mass would be important here). There are a lot of low age
clusters ($T<1$ Myr) in the figure because we arbitrarily start the
disruption process at $T=1$ Myr. This excess of low age clusters can be
ignored here because their numbers are arbitrary: we could start
cluster disruption earlier, for example. In reality, they would
represent embedded clusters still in the process of formation, before
any cloud disruption begins and before they move significantly from
their formation sites and get a chance to collide with anything.

Also shown in the top panel of Figure \ref{clusdest7} is the
distribution function of the ages, as in previous figures, but this
time for the mass interval $\Delta \log M/M_\odot=3$ to 5. For the high
formation rate (left panels), the age distribution starts somewhat flat
in the disruption era ($T>1$ Myr), as expected for a disruption rate
that increases linearly with cluster age. However, in making this
statement, we have to be careful to choose the right mass range because
the cluster mass function does not keep its initial $dN/dM\propto
M^{-2}$ form.  For the lower formation rate (middle panels), where the
model runs for a longer total time, the mass function changes so much
that there is no age range where the age distribution is flat.

The top right-hand panel of Figure \ref{clusdest7} shows the
distribution function of cluster masses in the low star formation rate
case for the time interval $\Delta \log T/{\rm Myr} = 2$ to 3. The
bottom right panel shows the mass function for the high formation rate
case and the time interval $\Delta \log T/{\rm Myr}=1$ to 2.  Both the
mass functions and the distributions of points in the $\log M-\log T$
plots illustrate the remark raised earlier for cases when the
disruption time is an increasing function of cluster mass: high-mass
clusters are not destroyed as rapidly as low-mass clusters in
proportion to their birth ratios, so the low-mass end of the mass
function gets depleted over time.  Such a turnover in the mass function
has not been observed for real disk clusters yet, but it could be below
the detection threshold.

The mass function for old clusters is
interesting because it resembles somewhat the peaked mass function of
globular clusters.  Above $M=10^6\;M_\odot$, the cluster mass is so
large that the disruption time in equation (\ref{dis2}) is longer than
the age, $T=10^3$ Myr. Then the clusters still have their initial
masses and the mass function is $dN/dM\propto M^{-2}$. Below
$M\sim10^5\;M_\odot$, the cluster disruption rate is very high and
clusters move almost directly downward on this plot, with a faster rate
for lower masses. This clears out the low mass clusters, giving the
observed peak mass.  As suggested elsewhere for globular cluster
models, the peak is approximately the mass where the disruption rate
equals the age. This mass increases over time.

\subsection{Dispersal of Hierarchical Stellar Groups}\label{disphier}

A third model of cluster disruption is relevant when the spatial
resolution of a star-forming region is not adequate to see the core
radius of a single bound cluster, which is several tenths of a parsec
for young clusters in the solar neighborhood (Testi, Palla, \& Natta
1999). What is observed instead is usually a collection of clusters and
associated free stars that is either marginally bound as a whole or
unbound in pieces (e.g., see images in Maiz-Apellaniz 2001; Bastian et
al. 2005b). In these cases, the separate pieces in the hierarchy may be
unbound even if each piece is self-bound. This means that over time,
the pieces can drift apart so what was once observed to be a single
massive cluster is later observed to be two or more lower-mass
clusters.

Also if clusters are born moderately bound, or if they become unbound
after gas removal, then their dispersal should be clumpy and not
smooth. Free expansion of unbound aggregates of stars produce mildly
bound sub-aggregates (Gerola, Carnevali, \& Salpeter 1983). Thus
cluster unbinding is always hierarchical in this sense.

Figure \ref{clusdest6} shows a model for this case in the right hand
panels (the left and center panels will be discussed in the next
subsection). Clusters were formed with an $M^{-2}$ mass function and
time step $dt$ like before, but now at each timestep for each cluster a
probability for fragmentation was evaluated:
\begin{equation}
P_{\rm frag}=\chi_3 dt/T.\end{equation} A random number between 0 and 1
was also generated for each timestep and cluster, and if for a
particular cluster the random number was less than $P_{\rm frag}$, then
that cluster was replaced by two clusters of lesser mass.

The masses used for the replacement clusters have a certain requirement
in order to give a uniform density on a $\log M-\log T$ plot. Suppose
there are 2 subclusters of masses $x_1M$ and $x_2M$ to replace the
chosen cluster of mass $M$.  Suppose also $\chi_3=2$ for the moment.
This means that the cluster fragmentation rate is twice $dt/T$, so in
all likelihood, two fragmentations will occur in a total time $T$ from
the current age $T$. Then each cluster has two chances of fragmenting
in a future time equal to its age.  Now we know from the required 45
degree angle of a cluster track in the $\log M-\log T$ plot (Sect.
\ref{evap}) that if a cluster age is to double, then its mass has to
decrease by a factor of 2. In the case of fragmentation, this means
that the total mass of the two fragments formed after a time equal to a
cluster age has to decrease by a factor of 2 from the total mass of the
single cluster before fragmentation. After the first fragmentation
event (in a future time equal to half the cluster age), there are two
clusters with total mass $(x_1+x_2)M$. After the second fragmentation
even (i.e., after the full time equal to the cluster age), there are
subcluster masses $x_1^2$, $x_1x_2$, $x_2x_1$ and $x_2^2$, because each
fragment from the first event breaks apart again into subfragments with
the same mass ratio. Thus the total mass after the second fragmentation
is $(x_1+x_2)^2M$. This has to equal half the original cluster mass,
$M/2$, so we have the constraint that $(x_1+x_2)^2=1/2$. If each
fragment has the same mass $(x_1=x_2)$, then $x_1=x_2=1/8^{0.5}=0.35$.
In general, for two fragments, $(x_1+x_2)^{\chi_3}=1/2$. For four
fragments in each fragmentation event,
$(x_1+x_2+x_3+x_4)^{\chi_3}=1/2$, and so on. For $N$ equal fragments
per fragmentation event, each has to have a mass equal to the fraction
$x=(0.5)^{1/\chi_3}/N$ of the cluster mass before fragmentation. The
case on the right in Figure \ref{clusdest6} has $\chi_3=2$ and $N=2$.
The maximum possible initial cluster mass in the distribution function
that is sampled during cluster formation is $10^8\;M_\odot$ to ensure
that clusters are still present in the $100-1000\;M_\odot$ range for
counting after many fragmentation events.

This fragmentation model does not preserve total cluster mass as the
total mass has to decrease by a factor of 2 for each doubling in age
(in order to give a uniform distribution on a $\log M-\log T$ plot).
The stars that do not remain in clusters drift into the field.

\subsection{Partial Cluster Disruption at Sudden Events}\label{partial}

Figure \ref{clusdest6} also shows two models where clusters do not
disappear completely, as they did in Figure \ref{clusdest}, nor do they
lose mass steadily as in Figure \ref{clusdest2}, but they suddenly lose
some fraction of their stars to the field and keep only the remaining
fraction in a bound clustered state.  On the left in the figure is a
case where the partial disruption probability in timestep $dt$ is
$2dt/T$ for cluster age $T$, and where each cluster chosen for partial
disruption (i.e., chosen by picking a random number and comparing it to
the probability, as above) has half of its mass removed. In the center
panels, the partial disruption probability is $4dt/T$ and each chosen
cluster has $1/4$ of its mass removed. The removed cluster stars are
assumed to go into the field where they do not contribute to the
observed cluster mass.

The mass fractions follow from the partial disruption rates as follows.
For a uniform distribution over $T$ on a $\log M-\log T$ diagram, we
need on average that half of each cluster remains after twice that
cluster's age. This puts clusters on a track in a $\log M-\log T$ plot
that has a 45 degree angle, and so it preserves the uniform
distribution in age $T$ for an $M^{-2}$ initial cluster mass function.
This means we can write the mass loss rate as
\begin{equation}
{{\Delta M}\over{\Delta T}} = -{{fM}\over {fT}}=-{M\over
T}\end{equation} as required if each loss event removes the fraction
$f$ of the cluster mass, and the mass loss time interval $\Delta T$ is
the fraction $f$ of the formation time interval, $dt$.  For the left
and center cases in Figure \ref{clusdest6}, $f=1/2$ and $1/4$,
respectively.

The left and center bottom panels in Figure \ref{clusdest6} also show
tracks for 5 separate clusters as they evolve with sudden partial
disruptions. As expected, the steps are two times bigger for the $f=2$
case, and there are half as many of them, compared to the $f=4$ case.
Still, all of the tracks have a average angle of 45 degrees in this
diagram.

\section{False Mass Loss by Peripheral Cluster Fading below a Surface Brightness
Limit}\label{false}

Clusters with a King (1962) profile have extended envelopes of stars
out to a tidal radius $R_t$.  If the surface brightness limit,
$I_{SB}$, is reached at a radius smaller than $R_t$, then the outer
part of the cluster may be missed and the mass determined from aperture
photometry may be too low. Cluster dimming makes the apparent loss of
mass increase over time.  We show here how this dimming affects the
distribution of clusters on the $\log M-\log T$ diagram.

The King (1962) profile of surface density in a cluster is
\begin{equation}
\Sigma(R)=k\left(\left[1+x\right]^{-1/2}-
\left[1+x_t\right]^{-1/2}\right)^2 \equiv
k\beta\left(R,R_c,R_t\right)\label{kingsb},\end{equation} and the
cumulative mass is
\begin{equation}
M(R)=\pi R_c^2 k \left[\ln\left(1+x\right) -4{{\left(1+
x\right)^{1/2}-1}\over{\left(1+x_t\right)^{1/2}}} + {{x}\over
{1+x_t}}\right] \equiv \pi R_c^2 k\gamma\left(R,R_c,R_t\right),
\label{kingmass}
\end{equation} where $x=\left(R/R_c\right)^2$ for radius $R$ and core
radius $R_c$, and for $x_t=\left(R_t/R_c\right)^2$; $k$ is a constant
determined from the total cluster mass $M(R_t)$ and core radius $R_c$
using equation (\ref{kingmass}). The tidal radius $R_t$ depends on the
external tidal field and total cluster mass. We assume the external
tidal field is constant (unlike the discussion in Sect. \ref{hier}),
and because total $M$ is about constant with time, $R_t$ is constant
also.

For a first set of models, a cluster is considered to have a constant
core radius with time and to dim uniformly with evolutionary effects at
the rate $\Psi_0(T/T_0)^{-\alpha}$ for initial light-to-mass ratio
$\Psi_0$; we use $T_0=1$ Myr for simplicity in normalization. Then the
outer detectible radius $R_d$ is given by the solution for $R$ in
equation (\ref{kingsb})
\begin{equation}
k\beta\left[R_d(T),R_c,R_t\right]\Psi_0(T/T_0)^{-\alpha}=I_{SB}.\label{kbeta}\end{equation}
The detectible mass follows from equation (\ref{kingmass}) but with
another modification from stellar evolution. In the single stellar
population models by Bruzual \& Charlot (2003), the mass of an initial
stellar population decreases with time approximately as
$(T/T_0)^{-0.078}$, as determined mostly by the loss of high mass stars
for the Chabrier IMF with solar metallicity. Thus we use a mass out to
the detectable radius $R_d(T)$ from the equation
\begin{equation}M(T)=\pi
R_c^2k(T/T_0)^{-0.078}\gamma\left[R_d(T),R_c,R_t\right].\label{kingmass2}
\end{equation}
The surface brightness limit, $I_{SB}$, is normalized to half of the
peak surface density in the cluster with lowest-mass, which is
$10\;M_\odot$; this is written as
$I_{SB}=0.5I_0\left(10\;M_\odot\right)$ in the figures, where $I_0$
follows from equation (\ref{kingsb}) with $R=0$.  In the simulations,
the cluster masses are chosen randomly from a $M^{-2}$ initial cluster
mass function, as before, and the constant $k$ for each cluster follows
from the total cluster mass, $R_c$, $R_t$, and $R=R_t$ according to
equation (\ref{kingmass}). Time is stepped along for each cluster to
follow the changing apparent cluster mass with age.

Figure \ref{clusdest8} shows the results on a $\log M - \log T$ plot
for three values of $\alpha$: 0.7, 1, and 1.3. As before, the top
panels show the density of plotted points (clusters) in equal intervals
of $\log T$ for masses between $\log M=1$ and 2. The density of points
is about constant over $\log T$ when $\alpha\sim1$. It increases with
$T$ when $\alpha<1$ (left panels) and decreases with $T$ when
$\alpha>1$ (right panels).  Colored curves show sample $M-T$ loci of
individual clusters.

These results may be understood from approximations to equations
\ref{kingsb} and \ref{kingmass} in the limit where the tidal radius
goes to infinity:
\begin{equation}
\Sigma(R)\sim k\left(1+x\right)^{-1}\end{equation}
\begin{equation}
M(R)\sim \pi R_c^2k\ln\left(1+x\right).\end{equation} As above, we take
the surface brightness limit as half ($=1/\eta$) the peak intensity
value for a $10\;M_\odot$ cluster. This peak intensity implies
$k\Psi_0=I_0(10\;M_\odot)$. Then the initial apparent radius for a
$10\;M_\odot$ cluster is given by $1+x=\eta$, and the initial apparent
radius for a cluster of mass $M$ is given by $1+x=\eta
\left(M/10\;M_\odot\right)$. For $y=M/10\;M\odot$, the limiting
observable radius after fading is given by
\begin{equation}
1+x=\eta y T^{-\alpha}\end{equation} for $T$ in units of $T_0=1$ Myr.
The observable mass at this time is
\begin{equation}
M_{obs}=\pi R_c^2 k y \ln \left(\eta y
T^{-\alpha}\right).
\end{equation}
Taking the derivative of $M_{obs}$ with respect to $T$ and rearranging,
we get
\begin{equation}
{{d\ln M_{obs}}\over{d\ln T}}={{-\alpha}\over{\ln\left(\eta y
T^{-\alpha} \right)}}. \end{equation} Previous sections had $d\ln
M/d\ln T=-\chi$, so now $\chi$ depends on $\alpha$, $\eta$, $M$
(through $y$), and $T$.  Figure \ref{clusdest8} shows the values of
$\ln M/d\ln T$ as the slopes of the curves in the bottom panels. They
have the same qualitative dependence as in this equation: larger
negative slope for larger $\alpha$ and $T$, and lower $M$.  The
logarithmic term requires $\eta y T^{-\alpha}>1$ for positive mass
values. The drop in the green curves in Figure \ref{clusdest8} occurs
when this term approaches 1.

The distribution in Figure \ref{clusdest8} differs from some of the
others in this paper in having a maximum detectible cluster mass that
increases with $\log T$, and having a minimum detectible cluster mass
that also increases with $\log T$. The first effect arises because
massive clusters are so bright that they stick high above the surface
brightness limit and their observable mass decreases very slowly at
first. The second effect arises because low mass clusters are quickly
lost below the surface brightness limit and drop off the diagram. We
also show in Figure \ref{clusdest8} some rising red lines, which
represent the slopes of the fading limits in each case. In a real
observation, only clusters brighter than a line parallel to this red
line can be observed at a certain magnitude limit.  This magnitude
limit could be confused with a surface brightness limit if the loss of
peripheral cluster stars is not recognized.

The increase of point density with $T$ for the most realistic case of
$\alpha=0.7$ suggests that cluster fading is not a good explanation for
the observed distribution on a $\log M - \log T$ plot. However, the
plotted density can be more constant if the core radius changes with
age in the right way. Returning to the simple expression for the King
profile, we add a dependence on core radius $R_c(T)$ that keeps the
mass constant,
\begin{equation}
\Sigma(R_c,T)=kyT^{-\alpha}(R_c[T]/R_0)^{-2}\left(1+x\right)^{-1}=k/\eta.\end{equation}
If $R_c/R_0\propto T^{0.5\left(1-\alpha\right)}$, then the apparent
radius is given by $1+x=\eta y/T$, and $d\ln M/d\ln T=-1/\ln\left(\eta
y/T\right)$, which is close to $\chi=1$ for intermediate values of $T$.

Figure \ref{clusdest8_expansion} shows examples of $\log M-\log T$
plots with $R_c\propto T^{0.5\left(1-\alpha\right)}$ for $\alpha=0.7$
and 1.3. The red rectangle shows the mass range where the point density
distributions are determined; the lower time limit to the rectangle is
where evolutionary effects are assumed to begin. The point density
distributions are flatter than without the $R_c$ variation in Figure
\ref{clusdest8}, as expected, but they fall off at high $T$ because of
the missing low mass clusters. The rising red line with a slope of 1
indicates the approximate lower boundary of these missing clusters (now
the slope is 1 because of the $R_c$ variation, whereas in Fig.
\ref{clusdest8}, it was $\alpha$). We can adjust this lower boundary by
varying the threshold surface density. The boundary is there because
the threshold is relatively high ($I_{SB}$ is half the initial central
surface density of the lowest mass cluster), so the low mass clusters
are lost from view quickly. We can keep more of these clusters if we
lower $I_{SB}$. This may be seen by comparing the left and middle
panels in Figure \ref{clusdest8_expansion}.

To determine how the lower boundary of missing clusters scales with
$I_{SB}$, we return to equations (\ref{kbeta}) and (\ref{kingmass2}).
This lower boundary is essentially where $R_d=0$ because then the
center of the brightest missing cluster is just at the surface
brightness limit. Equations (\ref{kingsb}) and (\ref{kbeta}) indicate
that at $R_d=0$, the limiting cluster has $k\sim I_{SB}(T/T_0)^\alpha$
when $x_t\sim\infty$. Also from the definition of $k$ in terms of $M$,
at the limiting mass $k=M/\left(\pi R_c^2 \gamma_t\right)$, where
$\gamma_t=\gamma(R_t,R_c,R_t)$. (In this expression, substitute
$M(T/T_0)^{0.078}$ for $M$ to account for stellar evolution mass loss).
As a result, the lowest detectable mass is given by
\begin{equation}
M_{min}\sim I_{SB}(T/T_0)^{\alpha} \pi R_c^2\gamma_t.\end{equation} We
see that $R_c\propto (T/T_0)^{0.5(1-\alpha)}$ makes this scale linearly
with $T/T_0$, as the red line indicates in Figure
\ref{clusdest8_expansion} (or scale with $(T/T_0)^\alpha$ when
$R_c=$constant, as in Fig. \ref{clusdest8}). We also see that
$M_{min}\propto I_{SB}$. Thus lowering the detection threshold lowers
the lower limit to detectable mass in direct proportion. The right two
panels of Figure \ref{clusdest8_expansion} show cases where $I_{SB}$
equals 0.2 times the central surface density of the lowest mass
cluster, instead of 0.5 times this value, which is on the left. The
points fill in the low-mass holes there a little bit, and the density
of points shown in the top panel is about constant for a longer range
in time.

The fading model does not have a constant density of points for all
mass ranges. There is a tendency to have more clusters per unit $\log
T$ at larger $T$. The mass range chosen has an approximately constant
point density because of competing effects by an increasing number from
the compression in $\log T$ space, and a decreasing number of low mass
clusters by surface brightness loss. Still, for some mass ranges, an
approximately constant density in a $\log M-\log T$ plot can result
from fading of the outer parts of clusters below the surface brightness
limit of the survey, given the usual model of stellar evolution with
$\alpha\sim0.7$, provided each cluster expands a little with age. The
ideal fit requires the clusters to expand as $T^{0.15}$, which
corresponds to a factor of 2 increase in core radius as the cluster
ages from 1 Myr to 100 Myr.  This expansion rate is consistent with
observation by Hwang \& Lee (2010).

\section{Cluster Selection Probabilities
in a Simulated Survey}\label{experiment}

The fundamental question addressed in this paper is why clusters appear
to be more and more missing from a survey as they age.  If the cluster
formation rate is constant over time and there is no disruption, then
there should be $X$ times more clusters in any interval of $\Delta \log
M$ and $\Delta \log T$ than in a comparable interval at a time $T/X$
before. In fact the number is about the same, so we ask where are the
missing clusters? If each cluster loses mass as $1/T$, then there are
no missing clusters from the $\Delta \log M \times \Delta \log T$ box:
the number is small because these are the same clusters that were in
the $\Delta \log M \times \Delta \log T$ box with higher mass, $XM$, at
the time $T/X$ before. There always were fewer clusters there, because
of the $M^{-2}$ mass function ($M^{-1}$ for equal log intervals of
$M$). If clusters do not lose mass slowly, but disrupt quickly, then
they are truly lost from the $\Delta \log M \times \Delta \log T$ box
and their stars are scattered into the field. Combinations of these two
processes can also account for cluster loss, as shown in Section
\ref{together}. We also considered other processes above, namely, cloud
collisions, hierarchical disassembly, partial sudden disruption, and
fading below the surface brightness detection limit.

In all cases, most of the stars lost from the missing clusters are
still present somewhere in the field. Thus we ask how likely is it to
see these stars with a more careful look. Pellerin et al. (2008)
consider this by searching resolved field stars for old cluster members
based on color-magnitude diagrams. As this method or others like it
become more advanced, it might be possible to reconstruct the $\log
M-\log T$ diagram using the total cluster mass including the field
stars formerly in the cluster. We predict that the density of points on
this plot will no longer be about constant with $\log T$, but will
increase in proportion to $T$, that is, $N(M,T)$ will lose its
$T^{-\chi}$ time dependence and show only the loss of mass from
supernovae and stellar winds.

To understand how observations of clusters can be susceptible to
detection limitations, we randomly placed template clusters from one
LMC field on another background LMC field and counted the proportion
that we could find by eye. Four clusters were used with absolute
magnitudes of $M_V=-10.3$, $-7.9$, $-7.5$, and $-5.4$, and ages of 19,
19, 30, and 16 Myr, respectively. After sky subtraction from the
cluster field, regions including the four clusters with radii of 25,
10, 15, and 6 pixels around them were cut out and embedded in a
$51\times51$ pixel image of zeros.  We then generated 20 lists of 100
random positions $x$ and $y$, excluding regions within 25 pixels of the
field edges. Cluster 1 was placed at the first 25 positions, cluster 2
at the next 25 positions, cluster 3 at the next 25 positions, and
cluster 4 at the last 25 positions. Four of these images were made for
the original clusters. This process was then repeated for 4 different
dimming factors, which were, including the original brightness, 1.0,
0.4, 0.1, 0.04, and 0.01. Figure \ref{clusterpix} shows the clusters
and their dimmed versions. In all, there were 20 fields in which to
search for clusters, and 100 clusters of varying brightness in each
field.

The embedding field was chosen to have a brightness gradient across it.
With random cluster positions, about half of the clusters ended up in
the bright part, and the other half ended up in the faint part. The
ratio of brightness is about 25:7. Figure \ref{a780} shows the field
before the addition of any template clusters.

One of us (DAH) went through each of the 20 images and marked the
positions of all things that looked like they were or could be
non-stellar, i.e. clusters. This was done the same way for all images.
She then compared the list of added cluster coordinates to the list of
identified objects and counted matches within 10 pixels radius as
detections of that added cluster. Some ``detections'' were made for
dimmed clusters that were really too faint to see (Fig.
\ref{clusterpix}), presumably because something else fuzzy was nearby.
Such uncertainty of cluster detection is inevitable at low brightness,
even in real surveys. The fraction of each cluster that was detected at
each dimming factor was determined, averaged over all 20 simulated
fields, with a distinction given to whether the cluster was found in
the bright part of the field or the faint part.

Figure \ref{clusdest_eyeball} shows the detection fractions, or
detection probabilities, as a function of cluster brightness, which is
defined to be the dimming factor multiplied by $10^{-0.4M_V}$. On the
left are the results plotted with one symbol for each cluster, dimming
factor, and background field brightness. On the right are 4 colored
lines that trace each of the 4 clusters along their dimming sequence (5
points per line). Evidently, the dependence of the detection
probability on cluster brightness is independent of which clusters,
dimming factors, and background fields were used.  The detection is in
fact a rather sharp threshold with a clear detection above the
threshold and a consistent miss below the threshold. This threshold
behavior explains why the lower limit to the mass in a $\log M - \log
T$ plot is relatively sharp and follows the fading trend with age $T$.
It does not explain how there can be a loss of clusters even above the
fading limit, since all clusters there should be detected.

The situation is about the same if we actually measure the simulated
cluster masses. To do this we ``observe'' the simulated cluster fields
in a realistic way, placing a circular aperture over each cluster at
its known position (whether or not it was found by the previous eye
examination), and subtracting a ``sky'' brightness taken from an
annulus 10 pixels wide at a radius equal to the unfaded cluster radius
plus 3 pixels. That is, the aperture radius was taken equal to the
original non-faded radius of the cluster: 25, 11, 16, and 7 pixels for
clusters 1, 2, 3, and 4, respectively.  Even clusters that were not
recovered in the eye examination tend to have some measured photometry
because there are field stars that get in the aperture. Some photometry
ends up as indefinite (``INDEF'') for various reasons. We then computed
the average of all non-INDEF magnitudes for each cluster in each half
of the image. The dispersion around the mean is taken to be the square
root of the ratio of the sum of the squares of the differences between
the measured magnitudes and the means, to the number measured. The only
noise was the noise already in the image to which the clusters were
added; no extra noise was added.

Figure \ref{eyeball_fading} shows the results of this fading
experiment. On the left are the instrumental magnitudes (subtract 1.9
for approximate calibrated magnitudes) determined for the clusters as a
function of the fading factors. Power law fits are shown for each
cluster and background field. The cluster magnitudes increase as their
intensities decrease with the fading factor. The slopes of the fits are
shown on the right in Figure \ref{eyeball_fading} versus the magnitudes
of the clusters before fading. The slopes average $-2.5$, which is the
value expected if the faded magnitude is fully recovered, i.e., without
loss from faint periphery. This result is consistent with that in
Figure \ref{clusdest_eyeball} in the sense that the cutoff between
observable clusters and unobservable clusters is sharp.  The magnitudes
of the clusters are correctly measured above the cutoff after
artificial fading.

We are puzzled why the experiment with real clusters does not reproduce
the expectation from the King profile discussed in Section \ref{false}.
Evidently the observed clusters that were cut out of the LMC,
artificially faded, and pasted on other fields for measurement had too
little mass at large radius to get significantly depleted after
dimming. The faded clusters appeared to shrink after dimming, and the
measurements were made on these smaller radii, but still the
luminosities came out correctly, i.e., proportional to the dimming
factors.  If surface brightness loss is not a factor in the $\log
M-\log T$ diagram, then the other cluster loss processes discussed in
this paper would have to be more important.

\section{Summary}\label{disc}

The observation for some galaxies of a nearly uniform density,
$N(M,T)$, of clusters within a specified mass interval on a $\log
M-\log T$ diagram places certain constraints on cluster disruption if
the cluster formation rate is about constant. Regardless of the
mechanism, it is necessary that about half of the current total cluster
mass be lost in twice the current cluster age. Or, as stated by Fall et
al. (2005), the number of clusters decreases by a factor of 10 for each
factor of 10 in age. Both of these relations give $N(M,T)\propto
T^{-1}$ for counting in linear intervals of $T$. This decrease factor
is not perfectly determined yet, as the observations tend to have poor
sampling statistics. It could be, for example, that the number
decreases by a factor of $\sim5$ for each factor of 10 in age (e.g.,
Fall, et al. 2007; Mora et al. 2009). Then $dN/dT\propto T^{-0.7}$. We
refer to the exponent here as the slope, $\chi$, of the number-age
relation, where the number refers to the number of clusters in a linear
age interval and a fixed mass interval. Similarly, $1-\chi$ is the
slope of the density-age relation, where density refers to the density
of points on a $\log M-\log T$ diagram.

In fact, there may not be a long-term steady decrease in the cluster
population at all, because the cluster formation rate is often not
known well enough to be certain of the cluster disruption rate (e.g.,
Bastian et al. 2009a). There are also direct (Hwang \& Lee 2010) and
indirect (Gieles \& Bastian 2008) indications that the density of
points on the $\log M-\log T$ diagram is not constant. However, if
there is a long-term, steady erosion of the cluster population, then
the reasons for this have to be determined. It cannot be the result of
either cloud collisional destruction or standard evaporation in a
time-invariant environment.

Here we considered cluster environments that change with time as a
result of prolonged cluster movement out of a star complex. We modeled
the resulting cluster populations in several ways:

(1) Rapid total disruption of each cluster with a collision probability
per time step, which is the same as a collision rate, inversely
proportional to the cluster's age (Sect. \ref{instant}). This gives
$\Delta M/\Delta t=-\chi_1 M/T$ for the cluster mass loss rate, because
$\Delta M=M$, the total cluster mass that is lost all at once, and
$\Delta t=T$, the cluster's age; $\chi_1$ is a constant that is equal
to the slope of the number-age relation. This model fits many of the
observations well if $\chi_1\sim0.7-1$, and it still fits the
observations if the cluster mass function has an upper mass cutoff. The
motion of a cluster in the $\log M - \log T$ diagram is purely
horizontal until the cluster disappears suddenly.  This model is
consistent with the motion of a cluster through a kpc-size cloud
complex with a Larson (1981) density-distance relation, i.e.,
$\rho\propto 1/R$, because then the density of sub-cloud collision
partners varies inversely with cluster age, and this makes the
collision rate vary inversely with age.

\begin{figure}
\epsscale{1.0} \plotone{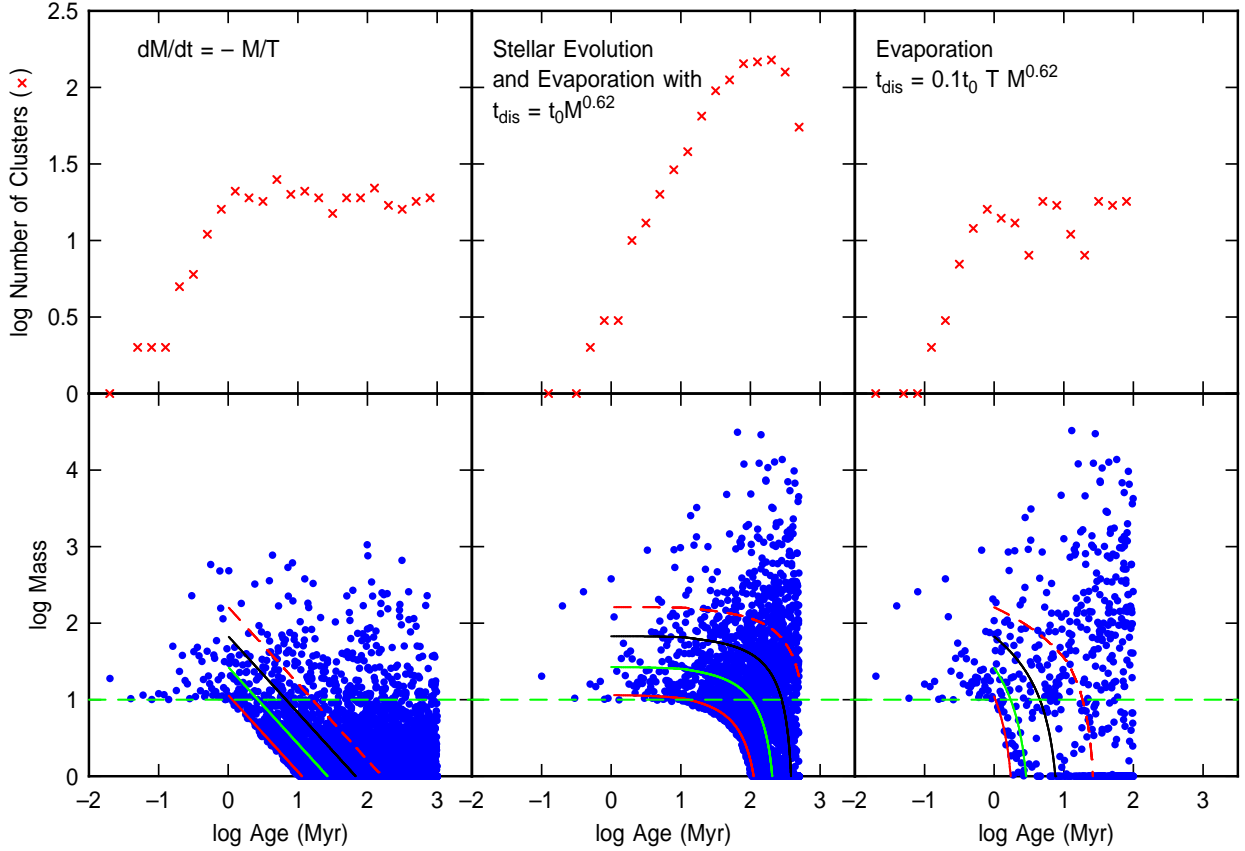} \caption{(Left) Model repeated from the
case $\chi_2=1$ in Figure 3, with lines in the bottom panel showing
tracks of individual clusters. (Middle) Model of cluster disruption
from Lamers, Anders, \& de Grijs (2006) that includes stellar evolution
mass loss and evaporation. The number of clusters in equal intervals of
$\log$ age increases sharply with age. The curves in the bottom figure
show individual cluster tracks. The horizontal parts of these tracks
are what cause the number per unit log age to increase. (Right)
Modified Lamers et al. model for the mass dependence of the disruption
time, with a linear age dependence. This age dependence causes the
curves in the bottom panel to have an average slope of about 45
degrees, like the panel on the left, and this is enough to make the
number of clusters per unit log age be about constant.
\label{clusdest2_lamersfit}}
\end{figure}

\begin{figure}
\epsscale{1.0} \plotone{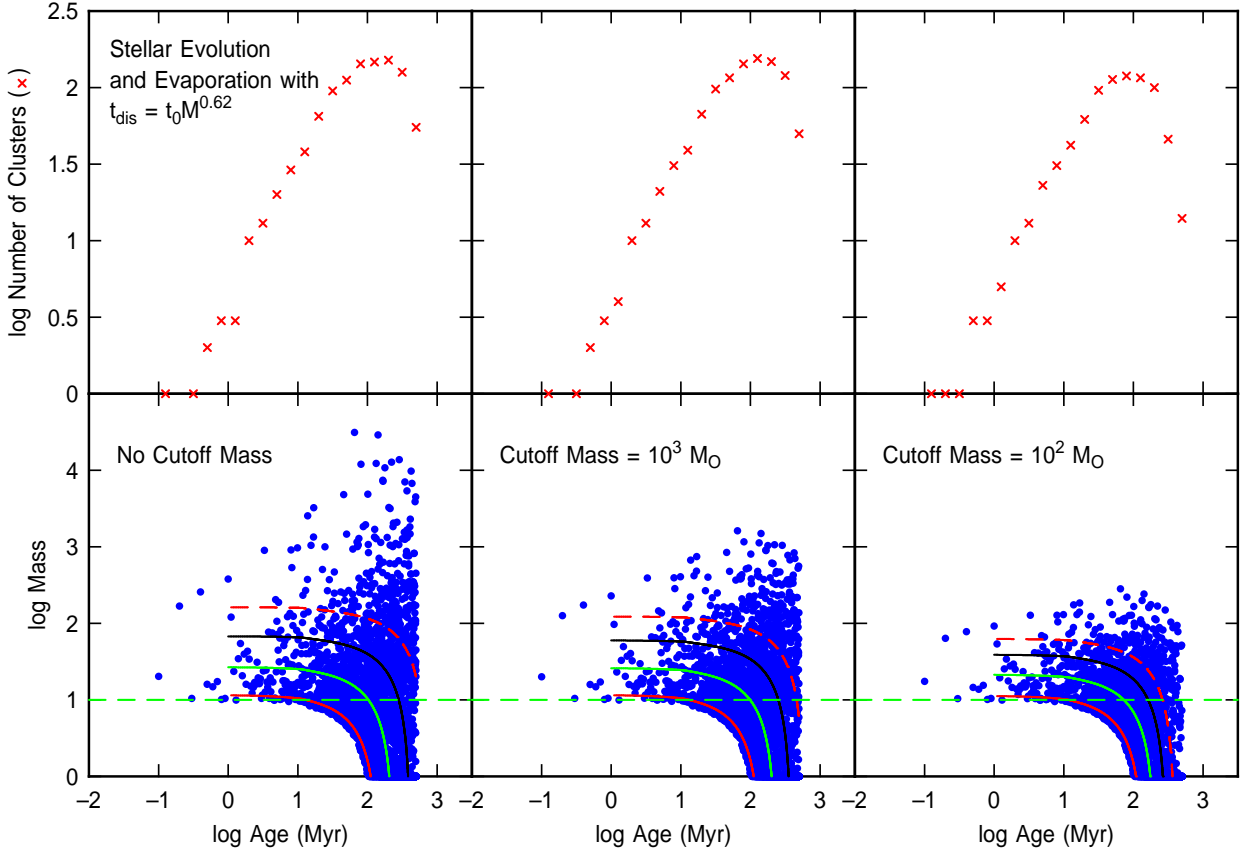} \caption{(Left) Same model as in the
middle of Figure 5, based on cluster disruption according to the
Lamers, Anders, \& de Grijs (2006) formulation.  (Middle) Model
repeated but with a Schechter mass function having a cutoff mass of
$10^3\;M_\odot$. (Right) Model repeated again but with a cutoff mass of
$10^2\;M_\odot$. The cutoff has little effect on the density-age
distribution (top panels) because the cluster evolutionary tracks are
nearly horizontal for most of a cluster's life.
\label{clusdest2_lamersfit4}}
\end{figure}

\begin{figure}
\epsscale{1.0} \plotone{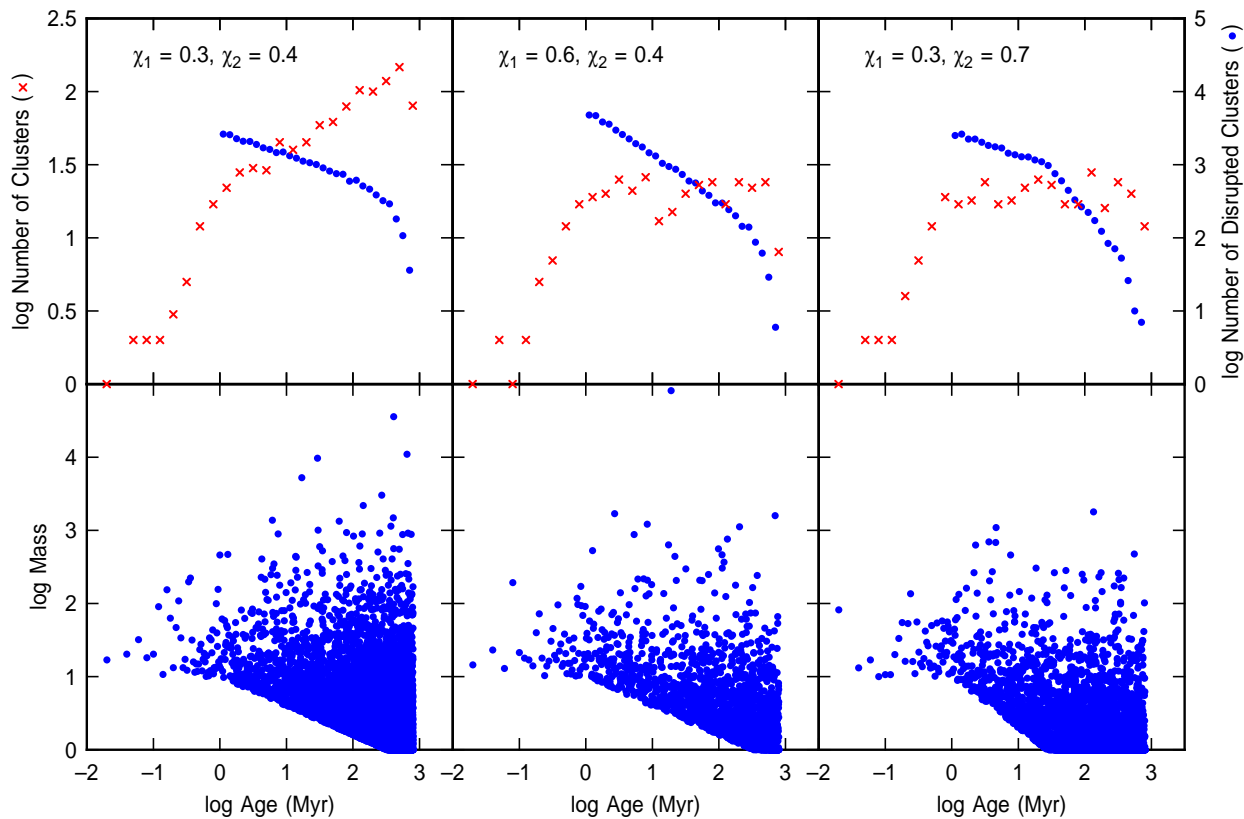} \caption{Models with continuous,
power-law mass loss for all clusters and sudden disruption for randomly
chosen clusters. The coefficients $\chi_2$ and $\chi_1$ were defined
previously for these two cases, respectively. The distribution in age
depends primarily on the sum, $\chi_1+\chi_2$. The crosses and dots in
the top panels are the distribution functions for the age of the
remaining clusters and the age of the disrupted clusters, as in Fig. 1.
\label{clusdest3}}
\end{figure}

\begin{figure}
\epsscale{1.0} \plotone{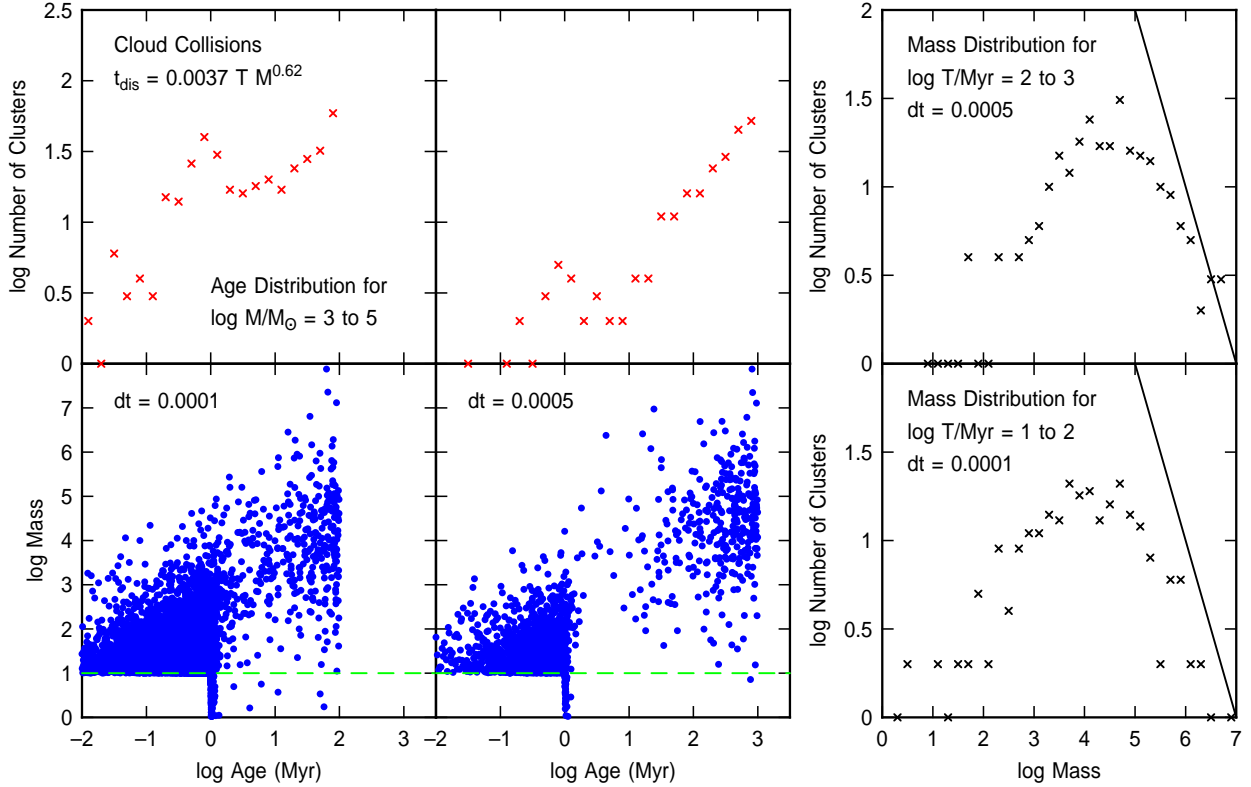} \caption{Models representing cluster
collisions with dense cloudy debris inside a star complex. The product
of the column density and average density of this debris is assumed to
vary inversely with cluster age as a result of the cluster's motion
away from its birthsite.  The disruption rate is very large for the
parameters chosen in this figure, so the tracks (not shown) of clusters
in the lower figure are nearly straight downward. The mass dependence
in the disruption time causes the most massive clusters to be
relatively little effected, and this leads to a change in the cluster
mass function from the initial power law to a kinked or peaked
function. Top panels on the left and center show the age distributions
for clusters in the mass range $\log M=3$ to 5.  (Left) High cluster
formation rate with a maximum cluster age of 100 Myr. (Center) Cluster
formation rate that is lower by a factor of 5, but with a maximum
cluster age of 1000 Myr. (Right) Mass distribution functions for the
two cases at ages in the range $\log T = 1$ to 2 for the bottom and 2
to 3 for the top. The mass functions have a
peak at the mass where the cluster age is about equal to its disruption
time. The difference in these two distributions is primarily the result
of cluster age. The solid line indicates the slope of the initial
cluster mass function. \label{clusdest7}}
\end{figure}

\begin{figure}
\epsscale{1.0} \plotone{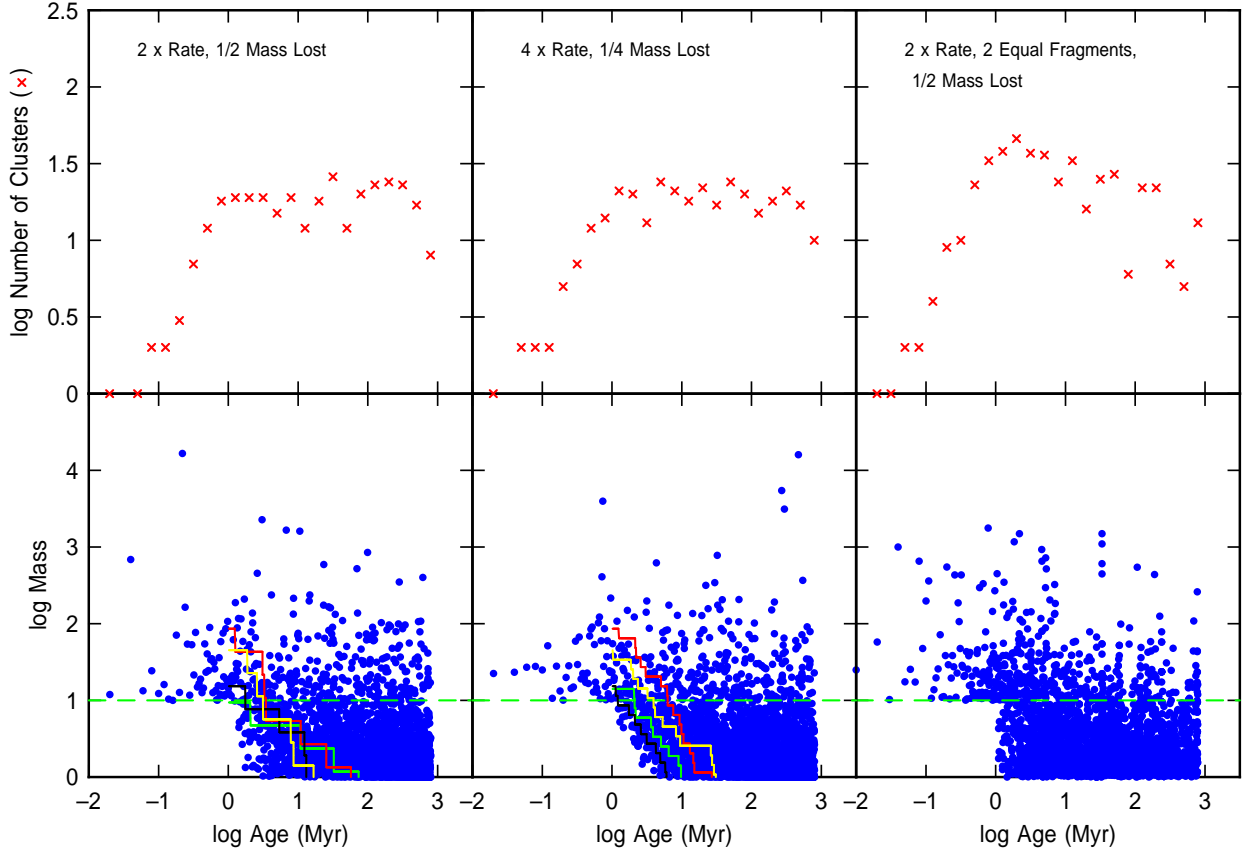} \caption{(Left and center) Models with
sudden but partial cluster disruption having fractional mass losses of
$f=1/2$ and $1/4$ at frequencies of 2 and 4 times the inverse age,
respectively. Jagged lines in the bottom panels show evolutionary
tracks for four sample clusters. (Right) Model with hierarchical
disassembly of clusters occurring at twice the frequency of the cluster
inverse age and with 2 equal mass fragments produced each
time.\label{clusdest6}}
\end{figure}

\begin{figure}
\epsscale{1.0} \plotone{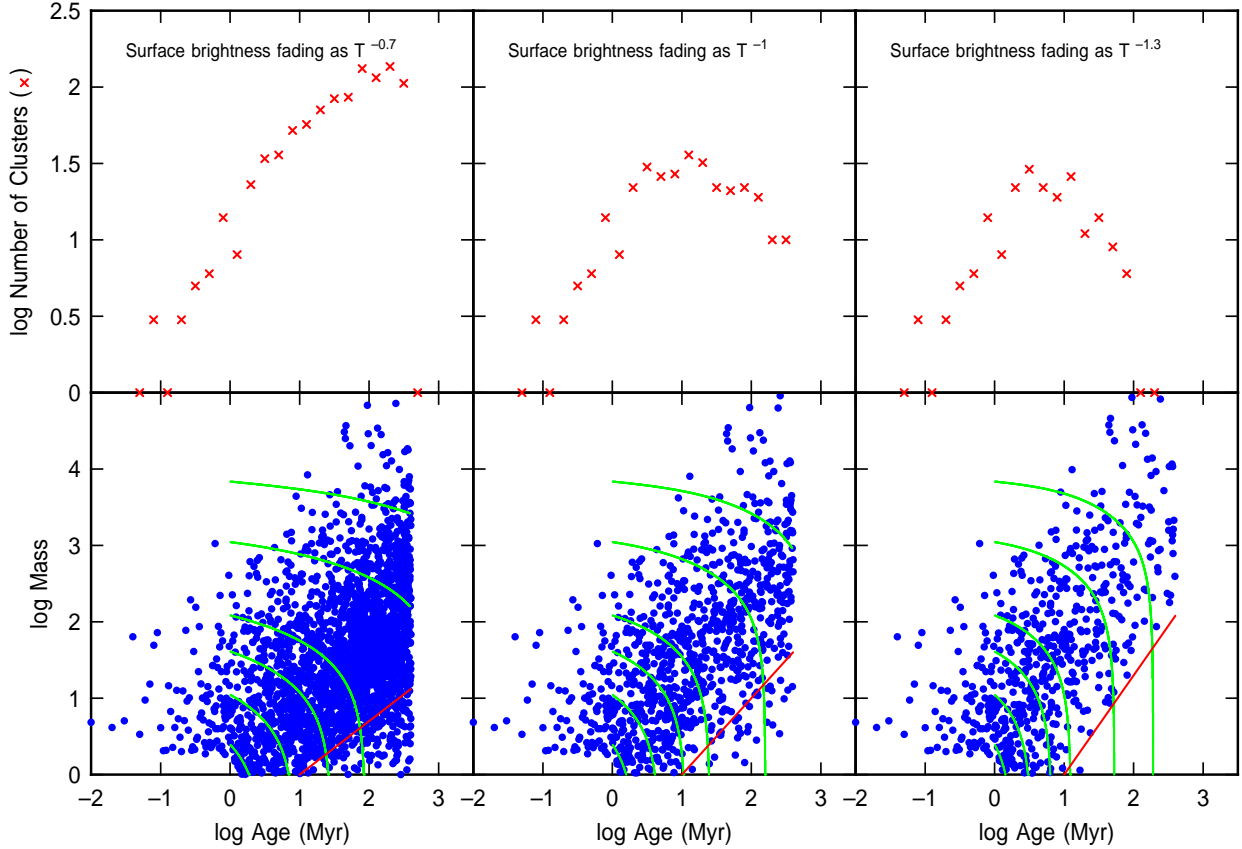} \caption{Models with cluster mass that
changes only slowly with stellar evolution (as $T^{-0.078}$) but with a
detectable mass including only the portion of the cluster brighter than
a fixed surface brightness limit. The mass in the cluster periphery is
progressively lost as the cluster fades. Three cases of fading are
considered: the realistic case with fading as $T^{-0.7}$ on the left, a
case with fading as $T^{-1}$, and another with fading as $T^{-1.3}$.
For all cases in this figure, the cluster core radius is taken to be
constant. Only the middle panel gives a reasonably flat distribution of
density on the $\log T$ axis for a fixed $\log M$ interval (taken
between $\log M=1$ and 2). The red line at the bottom of each panel
shows the slope of the fading function and also defines a sample lower
limit to cluster mass, below which clusters would be lost from view.
\label{clusdest8}}
\end{figure}

\begin{figure}
\epsscale{1.0} \plotone{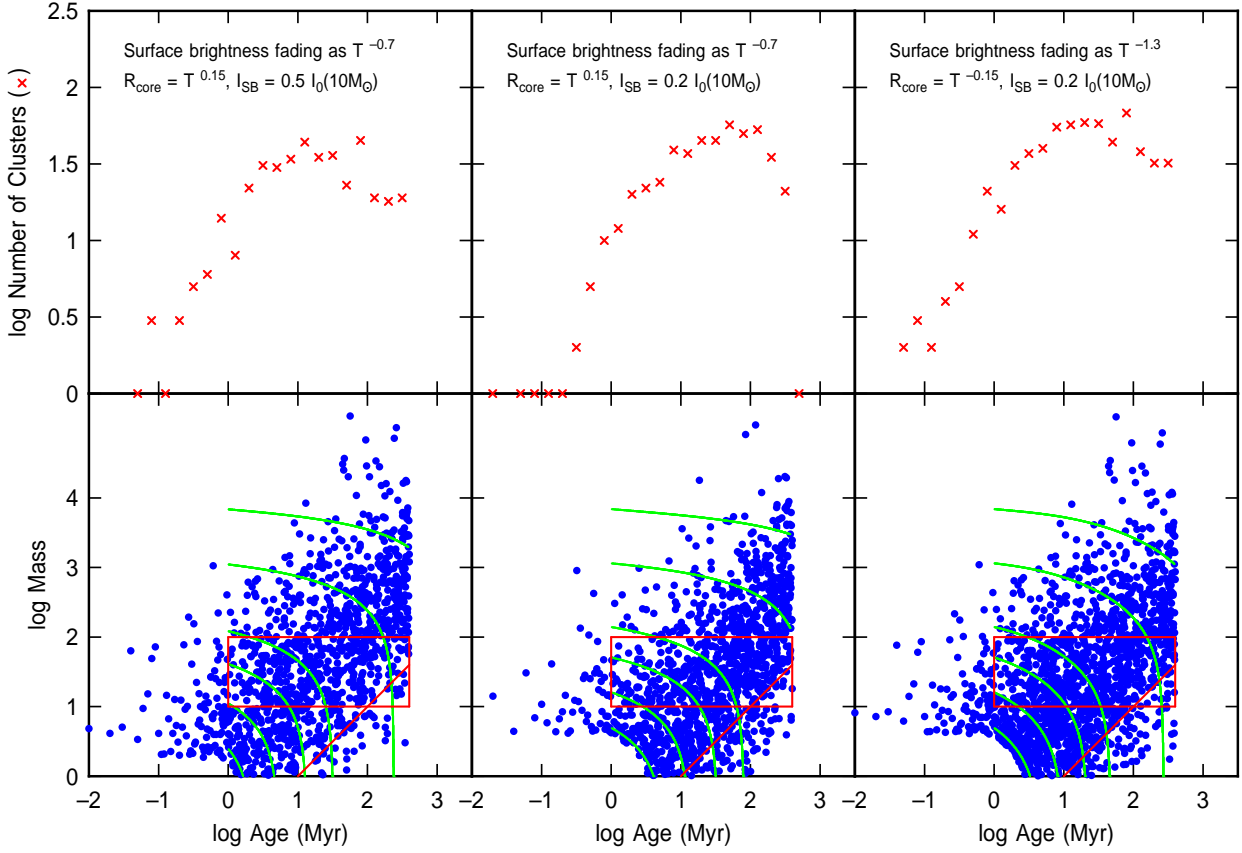} \caption{Models with cluster fading as
in Fig. 10, but now with time-changing cluster core radii, adjusted to
make the density of points nearly constant with $\log T$. The red
rectangle outlines the region used to determine this density, plotted
in the top panels. The red line has a slope of 1 in each case and
approximately defines the lower boundary of clusters. This lower
boundary depends on the fixed surface brightness limit. In the
left-hand panel, this limit is half the initial central surface
brightness of the lowest mass cluster. In the right two panels, it is
0.2 times this central brightness. The masses go lower in the center
and right panel because of this decrease in the surface brightness
threshold. \label{clusdest8_expansion}}
\end{figure}

\begin{figure}
\epsscale{1.0} \plotone{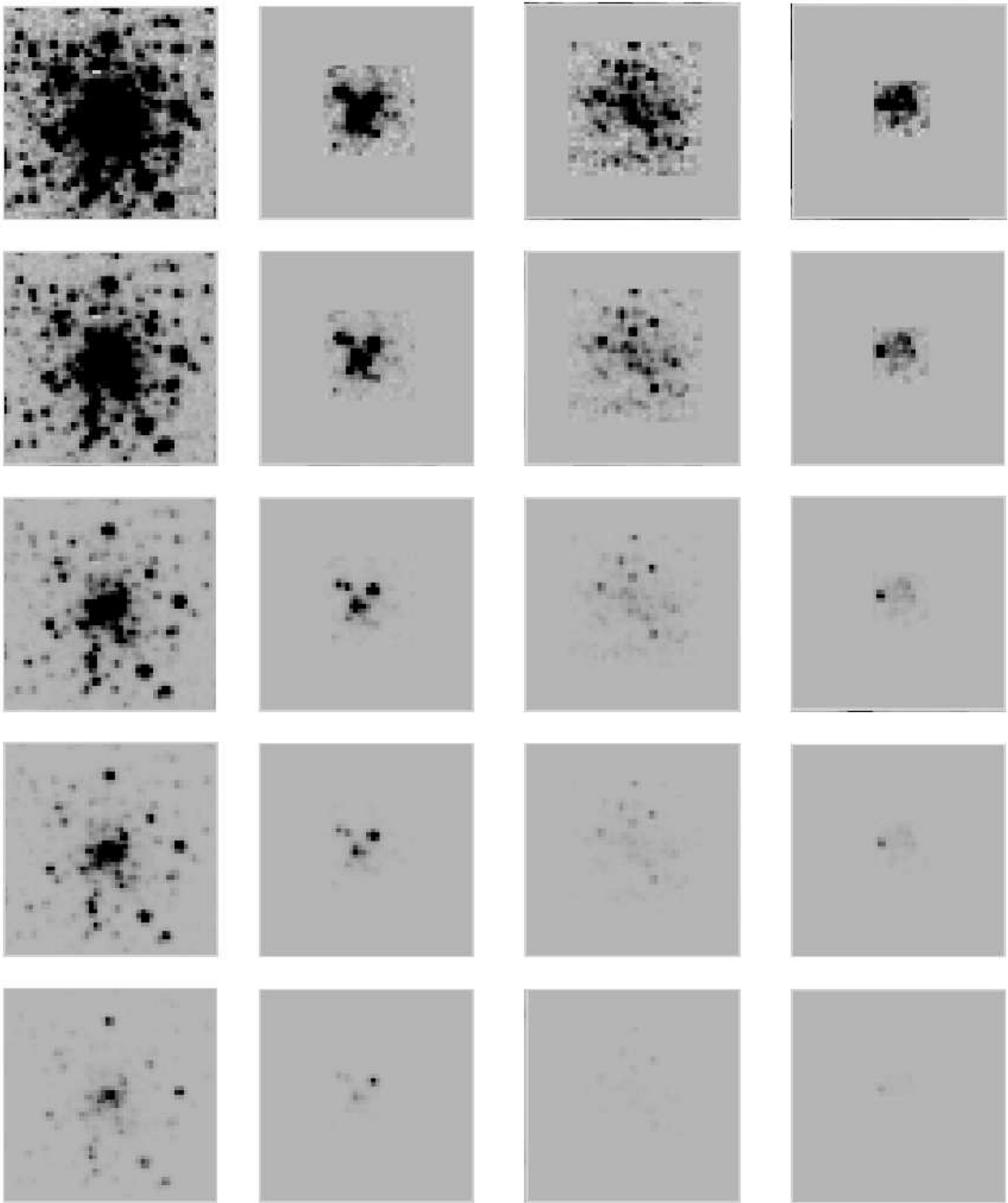} \caption{Four clusters, left to right,
with five levels of dimming, top to bottom, are shown here. These
cluster images were superposed on the field stars shown in Figure 13 in
order to measure cluster loss probability and apparent cluster masses
as a function of age. \label{clusterpix}}
\end{figure}

\begin{figure}
\epsscale{1.0} \plotone{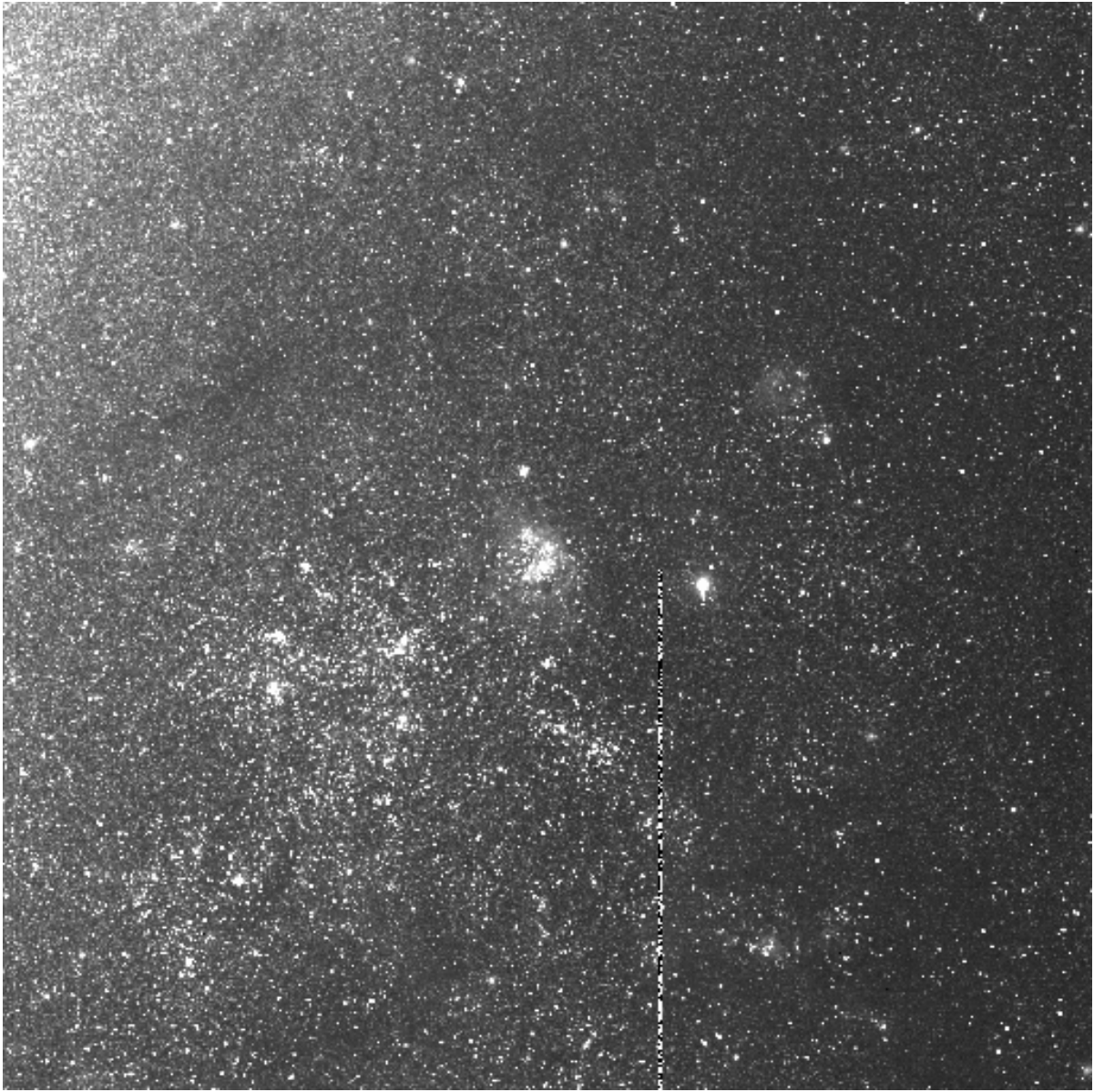} \caption{The LMC field used to place
the cluster images from Figure 12. The left side of the field has a
higher stellar background than the right side.\label{a780}}
\end{figure}

\begin{figure}
\epsscale{1.0} \plotone{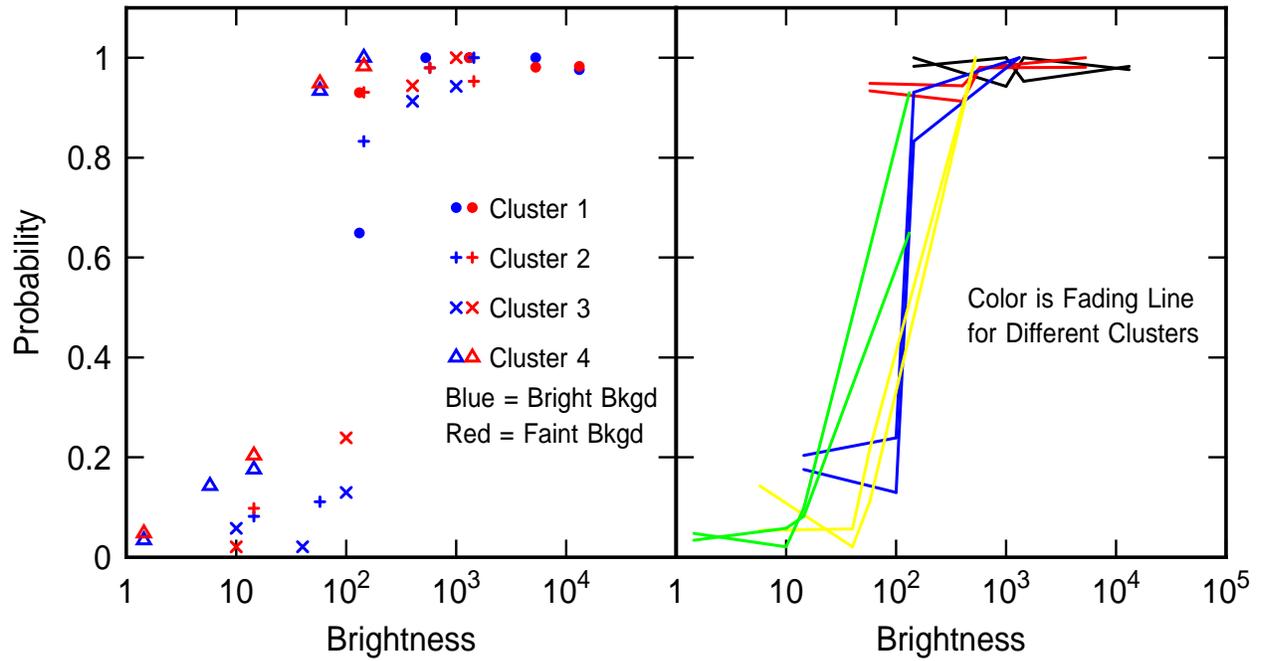} \caption{(left) The probability of
detecting an artificially dimmed cluster in a background field of stars
is shown versus the relative cluster brightness. (right) The detection
probabilities are shown as functions of cluster brightness for 4
clusters in two background fields, as indicated by separately colored
curves for each cluster.  \label{clusdest_eyeball}}
\end{figure}

\begin{figure}
\epsscale{1.0} \plotone{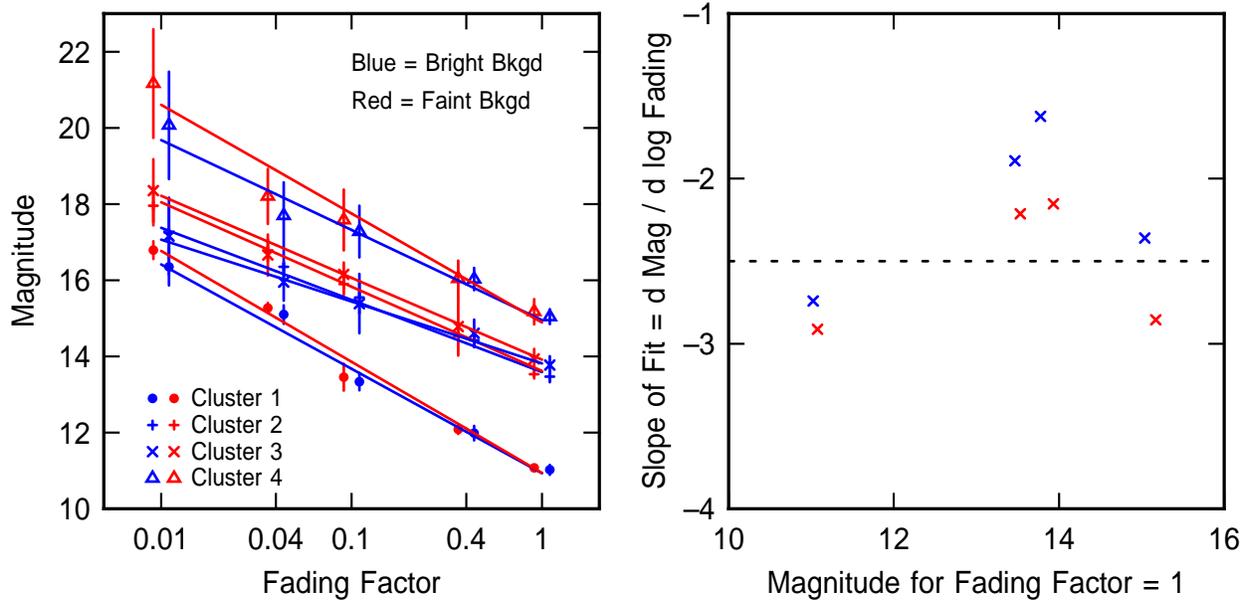} \caption{(left) The instrumental
magnitude of an artificially dimmed cluster in a background field of
stars as a function of the fading factor used to dim the cluster. These
are shown for clusters placed in the two stellar background levels
described in the text. Subtract approximately 1.9 mag for calibrated
magnitudes. (right) The slope of power law fits to $\delta \log
M/\delta \log T$ resulting from the magnitude dimming shown in the
right panel as a function of the initial cluster magnitude.
\label{eyeball_fading}}
\end{figure}

(2) Slow cluster mass loss at an instantaneous rate $dM/dT=-\chi_2 M/T$
(Sect. \ref{evap}). This is not the usual formula for thermal cluster
evaporation, which has either $dM/dt=$ constant in the standard model
(e.g., Spitzer 1987) or $dM/dt\propto M^{0.38}$ in the Lamers et al.
(2005) model. Any model like these two with a disruption time dependent
on mass has a distinct signature on a $\log M-\log T$ plot. The
evolutionary track of points on such a plot has an increasing downward
tilt with time, whereas each track has to have a constant downward
slope of magnitude $\chi$ (with an angle of ${\rm arctan} \chi$) in
order to give a density-age power law with slope of $1-\chi$ (provided
the initial cluster mass function is the usual power law, $dN/dM\propto
M^{-2}$). A modification of the Lamers et al. disruption time to make
it dependent on both age and mass gives somewhat better results (Fig.
\ref{clusdest2_lamersfit}: the ``modified Lamers model''), because the
initial evolutionary track is tilted downward and the rapidly falling
part at the end of the cluster's life can be below the detection limit.
A bigger problem with this slow-disruption model is that it is
incompatible with an upper mass cutoff in the cluster mass function.
The downward slope $\chi_2$ implies that massive old clusters with mass
$M_{max}$ and age $T_{max}$ had to have initial masses at time
$T_0\sim1$ Myr of $M_{max}\left(T_{max}/T_0\right)^\chi$, and this can
be a large value, much larger than a cutoff of around
$10^5-10^6\;M_\odot$. This model also has a problem with standard
evaporation in the star-complex environment because the varying tidal
density affects only the timescale for the mass loss rate, and not the
mass dependence. Evaporation would have predicted $dM/dT\propto 1/T$
for a tidal density that varies as $1/T^2$ (for isothermal cloud
structure in a star complex), and not the required $dM/dT\propto M/T$.
To get this extra mass factor in a slow-dispersal process, we would
have to assume that clusters disperse not by internal evaporation but
by repetitive cloud collisions, i.e., harassment. That is, we need a
model more like the first one to get the cluster mass in the numerator
of the mass-loss rate.

(3) A combination of the first two disruption mechanisms (Sect.
\ref{together}), giving the same total cluster mass loss rate: $dM/dt =
-\chi_1 M/T - \chi_2 M/T$ where the two terms are for rapid and slow
losses, respectively, and where $\chi_1+\chi_2$ is the slope of the
number-age relation. This model also agrees with observations to the
same degree as the first two models, and it is more likely than either
alone because clusters are expected to both lose mass slowly by
themselves and disperse suddenly during collisions.

(4) Rapid disruption by cloud collisions (Sect. \ref{coll}), as in
Section \ref{instant}, but using a disruption time $t_{dis}$ from
Gieles et al. (2006a). This disruption time is essentially the same as
in the Lamers et al. (2005) model for evaporation, but the coefficient
for $t_{dis}$ is smaller in the Gieles et al. model (i.e., there is
faster disruption by collisions than evaporation). To consider
collisions in a star complex environment, we modified $t_{dis}$ to have
a smaller initial numerical coefficient and we gave it a linear age
dependence. This change follows from the assumption that clusters are
born in a dense environment where collisions with cloud pieces are
frequent at first, and then the clusters drift into a lower density
environment where cloud fragments are less common. The results showed a
nearly constant density distribution in a $\log M-\log T$ plot, similar
to many observations. However, the mass distribution function changed
so much over time by the selective loss of low mass clusters that the
range of age giving the standard mass function $dN/M\propto M^{-2}$ was
limited. Because mass functions at intermediate age with low mass
turnovers are not observed yet, this model works only if the low mass
turnover is below the detection limit.

The model is interesting nevertheless because it suggests a way to make
peaked cluster mass functions for old clusters, similar to the mass
function for halo globular clusters. If this model is in fact
responsible for the globular cluster mass function, then most of the
cluster dispersal would have had to occur early on, in the disk
environment where the clusters formed. Once they are in the halo,
collisions with other objects, particularly with a $1/T$ rate, would be
relatively infrequent.  It is not known when the globular cluster mass
function first had its log-normal form, but it could have been very
early, with no change in shape from subsequent evaporation (e.g., see
models in Vesperini [1998, 2000] that show no time evolution of a
log-normal mass function in typical halo environments).  In this case,
globular clusters could have formed with the usual $1/M^2$ mass
function in the disk of a young (redshift $z\sim10$) galaxy and then
dispersed over the next 0.1 Gyr by cloud collisions in star complex
environments. This would have formed the log-normal mass function very
early in the life of the clusters.  Early galaxy collisions could have
then dispersed these clusters into the young galaxy halos, or minor
mergers of small cluster-forming galaxies with bigger galaxies could
have accumulated these clusters into the bigger galaxies' halos.

(5) Hierarchical cluster disassembly into $N$ pieces at each of a
series of disruption events (Sect. \ref{disphier}), with an event rate
$\chi_3/T$ and a summed mass fraction for the pieces equal to
$0.5^{1/\chi_3}$. Equal mass fragments would therefore need to have the
fraction $0.5^{1/\chi_3}/N$ of the remaining cluster mass at each
fragmentation event. The rest of the cluster goes into the field. This
model is reasonable considering that rapid cluster disassembly could
lead to fragmented pieces, and considering that poor resolution of
distant clusters could blend together initially unbound pieces which
then drift apart. Cluster formation is hierarchical in any case, so
there is some aspect of cluster disassembly that should be hierarchical
too, especially for very young, incompletely mixed, clusters.

(6) Rapid partial disruption giving a mass loss rate $\Delta M/\Delta T
=-fM/fT=-M/T$ for $f<1$. Here, partial disruption of the fraction $f$
of a cluster's mass occurs quickly when it happens, and the rate at
which is happens is $1/fT$ for cluster age $T$. The disrupted fraction
of the cluster's mass, $fM$, goes into the field. This model is a
variation of the first model summarized above, but is more flexible in
that it allows for partial disruption during a collision.

(7) Apparent loss of cluster mass by fading of a King-profile periphery
below the surface brightness limit of the survey. This gives a constant
cluster density on a $\log M - \log T$ plot for the standard fading
rate if cluster core radii expand slightly with age, as
$T^{0.5(1-\alpha)}$. In this notation, cluster luminosities fade with
age as $T^{-\alpha}$ in the absence of evaporation or disruption.
Stellar population modeling suggests that $\alpha\sim0.7$ so the core
radii would have to grow as $T^{0.15}$. Hwang \& Lee (2010) observe
this growth rate for clusters in M51. This model is a reasonable
explanation for cluster evolution on a $\log M - \log T$ diagram even
if there is no cluster disruption at all. Fading alone can explain the
distribution of cluster positions on this diagram. This implies that a
combination of cluster disruption by fading and by collisions or
harassment in a star complex environment can explain the observations.
Fading is inevitable, so perhaps this is the most reasonable situation
provided the peripheral mass in a cluster is unobservable below the
surface brightness limit of a survey.

In addition, we examined observations of clusters placed in bright and
faint fields with various degrees of artificial dimming in order to
determine the loss probability and apparent mass as a function of
brightness.  A cluster was lost suddenly from a field of view when its
magnitude dimmed below a certain value, thereby explaining the sharp
lower cutoff to observable cluster mass as a function of age. However,
the clusters that were observed above this limit had measured masses
that were correct for their dimming factors. This is contrary to our
expectations from the King-profile modeling, and suggests that the
artificially dimmed clusters had edges that were sharper than a King
profile. Perhaps they already lost mass below the surface brightness
limit before they were clipped and moved to other fields for the
simulated survey.

All of the successful cases have the property that the disruption or
mass loss timescales increase linearly with cluster age. Such an
increase is not part of any current cluster disruption model.  We
suggested a new model in which most cluster disruption occurs in the
extended dense and clumpy region surrounding the cluster's birthsite
(Sect. \ref{hier}). In a typical star complex, each cluster should
experience a time-changing tidal field and a time-changing density of
collision partners as it drifts and the cloud complex disperses. It is
possible that the basic disruption timescale then increases somewhat
smoothly with cluster age. In an alternative model, cluster loss by
fading gets its power law relation between detected mass and age from a
King-like profile for cluster surface density.

\acknowledgements We are grateful to Mark Gieles, S. Michael Fall, and
Rupali Chandar for providing useful references on the $\log M-\log T$
diagram. We are also grateful to Mark Gieles for suggesting we consider
a cutoff mass and for finding several small errors in early versions of
the manuscript. Funding for this research was provided by NASA to DAH
through grant NASA-GALEX NNX08AU57G. Funding to BGE was provided by NSF
grant AST-0707426.

\clearpage

\clearpage

\clearpage

\clearpage

\clearpage

\clearpage

\clearpage

\clearpage

\clearpage

\clearpage

\clearpage

\clearpage

\clearpage

\clearpage

\clearpage

\end{document}